\documentclass[useAMS,usenatbib]{mn2e}
\usepackage{aas_macros}
\usepackage{graphicx}	%
\usepackage{amsmath}	%
\usepackage{amssymb}	%
\usepackage{multicol}	%
\usepackage{bm}		%
\usepackage{pdflscape}	%
\usepackage{hyperref}
\usepackage{ulem} %
\usepackage{lineno} %

\newcommand{\RNum}[1]{\uppercase\expandafter{\romannumeral #1\relax}}

\newcommand{\redmagic}{\rm redMaGiC}
\usepackage[usenames]{color}
\definecolor{purple}{RGB}{150,0,200}
\definecolor{darkgreen}{rgb}{0.0, 0.5, 0.0}

\newcommand{\avg}[1]{\left\langle #1 \right\rangle}

\newcommand{\Omegam}{\Omega_{{\rm m}}}

\newcommand{\redmapper}{{\rm redMaPPer}}
\newcommand{\buzzard}{{\rm Buzzard}}

\newcommand{\buzzarda}{{\rm BuzzA}}
\newcommand{\buzzardb}{{\rm BuzzB}}
\newcommand{\buzzardc}{{\rm BuzzC}}

\defcitealias{DES_cluster_cosmology}{DES2020}
\defcitealias{DESY1KP}{DESY1KP}

\usepackage[T1]{fontenc}
\usepackage{ae,aecompl}

\usepackage{eso-pic}%

\AddToShipoutPictureBG*{%
  \AtPageUpperLeft{%
    \hspace{0.75\paperwidth}%
    \raisebox{-3.5\baselineskip}{%
      \makebox[0pt][l]{\textnormal{DES-2020-0560}}
}}}%

\AddToShipoutPictureBG*{%
  \AtPageUpperLeft{%
    \hspace{0.75\paperwidth}%
    \raisebox{-4.5\baselineskip}{%
      \makebox[0pt][l]{\textnormal{FERMILAB-PUB-20-329-AE}}
}}}%

\usepackage{txfonts}

\title[4x2pt+N Simulation]{Combination of cluster number counts and two-point correlations: Validation on Mock Dark Energy Survey}
\author[To\&Krause et al.]{
\parbox{\textwidth}{
\Large
Chun-Hao To,$^{1,2,3}$\thanks{Corresponding author: \href{mailto:chto@stanford.edu}{chto@stanford.edu}}
Elisabeth Krause,$^{4,5}$\thanks{Corresponding author: \href{mailto:krausee@arizona.edu}{krausee@arizona.edu}}
Eduardo Rozo,$^{5}$
Hao-Yi Wu,$^{6,7}$
Daniel Gruen,$^{2,3}$\\
Joseph DeRose,$^{8,9}$
Eli Rykoff,$^{2,3}$
Risa H. Wechsler,$^{1,2,3}$
Matthew Becker,$^{10}$
Matteo Costanzi, $^{11,12}$ \\
Tim Eifler,$^{4}$
Maria Elidaiana da Silva Pereira, $^{13}$and
Nickolas Kokron $^{1,2,3}$
\begin{center} (DES Collaboration) \end{center}
}
\vspace{0.4cm}
\\
\parbox{\textwidth}{
$^{1}$ Department of Physics, Stanford University, 382 Via Pueblo Mall, Stanford, CA 94305, USA\\
$^{2}$ Kavli Institute for Particle Astrophysics \& Cosmology, P. O. Box 2450, Stanford University, Stanford, CA 94305, USA\\
$^{3}$ SLAC National Accelerator Laboratory, Menlo Park, CA 94025, USA\\
$^{4}$ Department of Astronomy/Steward Observatory, University of Arizona, 933 North Cherry Avenue, Tucson, AZ 85721-0065, USA\\
$^{5}$ Department of Physics, University of Arizona, Tucson, AZ 85721, USA\\
$^{6}$ Center for Cosmology and Astro-Particle Physics, The Ohio State University, Columbus, OH 43210, USA\\
$^{7}$ Department of Physics, Boise State University, Boise, ID 83725, USA\\
$^{8}$ Department of Astronomy, University of California, Berkeley,  501 Campbell Hall, Berkeley, CA 94720, USA\\
$^{9}$ Santa Cruz Institute for Particle Physics, Santa Cruz, CA 95064, USA\\
$^{10}$ Argonne National Laboratory, 9700 South Cass Avenue, Lemont, IL 60439, USA \\
$^{11}$ INAF-Osservatorio Astronomico di Trieste, via G. B. Tiepolo 11, I-34143 Trieste, Italy\\
$^{12}$ Institute for Fundamental Physics of the Universe, Via Beirut 2, 34014 Trieste, Italy\\
$^{13}$ Brandeis University, Physics Department, 415 South
Street, Waltham MA 02453\\
}
}

\begin{document}
\label{firstpage}
\pagerange{\pageref{firstpage}--\pageref{lastpage}}
\maketitle

\begin{abstract}
We present a method of combining cluster abundances and large-scale two-point correlations, namely galaxy clustering, galaxy--cluster cross-correlations, cluster auto-correlations, and cluster lensing. This data vector yields comparable cosmological constraints to traditional analyses that rely on small-scale cluster lensing for mass calibration. We use cosmological survey simulations designed to resemble the Dark Energy Survey Year One (DES-Y1) data to validate the analytical covariance matrix and the parameter inferences. The posterior distribution from the analysis of simulations is statistically consistent with the absence of systematic biases detectable at the precision of the DES Y1 experiment.
We compare the $\chi^2$ values in simulations to their expectation and find no significant difference.
The robustness of our results against a variety of systematic effects is verified using a simulated likelihood analysis of a Dark Energy Survey Year 1-like data vectors. This work presents the first-ever end-to-end validation of a cluster abundance cosmological analysis on galaxy catalog-level simulations.
\end{abstract}

\setcounter{footnote}{1}

\section{Introduction}

The simple cosmological model of a vacuum dark energy and cold dark matter ($\Lambda$CDM) is able to describe a variety of observations from  the high- to low-redshift universe. 
Despite its success, the two pillars of this model, dark energy and cold dark matter, lack a fundamental theory to connect to the Standard Model of particle physics. Without a compelling candidate for such a theory, one way to test the $\Lambda$CDM paradigm is by comparing its predictions to precise measurements of %
both the growth of structure and the expansion history of the universe over the past several Gyrs, %
when dark energy dominates the total energy budget of the universe. %

Commonly used probes of growth and/or cosmic expansion include Type-Ia supernovae, galaxy clustering, weak gravitational lensing, redshift-space distortions, and the abundance of galaxy clusters (see e.g. \citealt{2013PhR...530...87W} for a review).
A large body of work has shown that the combination of these different probes is particularly powerful. For example, \citet[][\citetalias{DESY1KP} hereafter]{DESY1KP} %
combines three two-point correlation functions --- galaxy clustering, galaxy--galaxy lensing, and cosmic shear --- resulting in a precise constraint on the growth of the structure. Similar analyses have also been carried out for the Kilo-Degree Survey (KiDS, \citealt{KIDS3x2pt1, KIDS3x2pt2}). %
In this work, we extend this type of analysis by incorporating cluster abundances and cluster-based large-scale-structure statistics into the data vector of the combined probe analysis. 

Galaxy clusters form at peaks of the primordial matter density field, and their space density over time reflects the gravitational growth of  the coupled fluctuations of dark matter and baryons. 
 As such, the abundance and %
spatial distribution of galaxy clusters are sensitive to the growth of structure and the expansion history of the universe (see e.g. \citealt{2011ARA&A..49..409A} for a review). Due to their independent information, different systematic uncertainties, and different degeneracies, it is expected that the combination of cluster statistics with other cosmological probes will yield 
cosmological constraints that are both more precise and more robust \citep[e.g.][]{Takada&Bridle, Ogori,2014PhRvD..90l3523S,cosmolike2016, 2016JCAP...08..005L, 2020MNRAS.491.3061S,Andrina}.
Critically, however, despite the extensive theoretical work on this front, no implementation of these techniques have been validated on realistic cosmological survey simulations, nor applied to data.

This paper presents an essential step towards accurate cosmological parameter constraints from combined cluster statistics. %
We develop %
a model and covariance matrix to combine cluster information with two-point correlation functions, including galaxy clustering, galaxy--cluster cross-correlation, cluster clustering, and cluster lensing.  We validate the applicability of this model for a Dark Energy Survey Year 1-like experiment, which comprises 1321 $\rm{deg}^2$ area of the sky in five broadband filters, $g$,$r$,$i$,$z$,$Y$. In this study, we consider cluster samples built using the red sequence Matched-filter Probabilistic Percolation cluster finder algorithm (\redmapper{}; \citealt{Redmapper1}) and galaxy samples built using the automated algorithm for selecting Luminous Red Galaxies (\redmagic{}; \citealt*{Redmagic}).

Much of the information on structure growth available in current surveys lies beyond the regime where the theoretical modeling is perturbative, making a theoretical prediction of the observations challenging. The theory is challenged further when using galaxies and galaxy clusters as dark matter tracers, since such analyses %
require a sufficient understanding of their statistical connection to the dark matter. 
Moreover, the overdensities of galaxies and galaxy clusters can be subject to significant systematic biases due to observational constraints.
For galaxy clusters in particular, it is known that photometrically selected samples suffer from projection effects: massive dark matter halos are more easily identified as galaxy clusters when their own galaxy overdensities are enhanced in the plane of the sky by the projections of unassociated galaxies along the line of sight \citep*{Matteoprojection, Tomomi}.  This selection effect biases both the observed galaxy and matter overdensities about the selected galaxy clusters relative to randomly selected halos of the same mass.  A similar argument leads one to conclude that orientation biases, the major axes of detected clusters aligned with the line of sight, must also be present in photometrically selected cluster samples \citep{Heidiorientation}. Without a proper model of these systematics, the cosmological constraints from cluster abundances would be biased \citep[][\citetalias{DES_cluster_cosmology} hereafter]{DES_cluster_cosmology}.
The existence of important observational systematic biases implies that robust cosmological analyses that rely on galaxies and galaxy clusters should be tested on simulated data sets that explicitly incorporate as many of these systematics as possible \citep[see, e.g.][for an example applied to the DES 3$\times$2 point measurement]{Niall}. Such simulations are intended to provide plausible realizations of a given cosmology, allowing one to test the robustness of the analysis against both theoretical and observational systematics. This approach is particularly powerful when one considers the essentials of a blind analysis of survey data: %
simulations allow us to finalize analysis choices on %
simulated data sets prior to applying the method to the real data. 

However, there is an important caveat to this approach. No simulation is perfect.
When analyzing synthetic data, it can be difficult to disentangle biases in cosmological parameters coming from flaws in the analysis from those due to differences between simulations and data. The latter is often due to uncertainties in the underlying galaxy population or in the models for how galaxies trace the underlying dark matter density field \citep{Risa2018}. %
In this paper, we take a conservative approach and use three sets of cosmological simulations, each populated with galaxies using different assignment schemes,
to develop and test our theoretical model. 
Specifically, we will demonstrate that the theoretical model developed in this work is capable of correctly recovering the underlying cosmological parameters of several simulated data sets irrespective of the details of the galaxy population model.

This work considers four tracers that can be measured by an imaging survey: the abundance of galaxy clusters, the spatial distributions of galaxies and galaxy clusters, and the lensing shear field ($\gamma$).
These tracers are related to fluctuations of the matter density field, making them sensitive to the growth of structure in the universe. %
Because halos that host galaxy clusters form from rare peaks in the matter density field,
their abundance (the halo mass function) is highly sensitive to the the amplitude of matter fluctuations in the universe. By binning galaxy clusters based on an observable proxy for halo mass (such as the richness of \redmapper{} clusters), one may simultaneously calibrate the relation between this observational proxy and halo mass, as well as the underlying cosmological parameters.
Turning to the spatial distribution of galaxies and clusters, the main challenge to our ability to extract cosmological information from the corresponding correlation function is that both galaxies and clusters are biased tracers of the matter density field.
Fortunately, on sufficiently large scales, their overdensities are 
simply proportional to the matter fluctuations.
Even then, however, the amplitude of matter fluctuations are degenerate with the galaxy and cluster bias.

The above degeneracy can be broken using weak gravitational lensing.
Weak gravitational lensing shear is the coherent distortion of the shapes of distant galaxies (often called source galaxies) due to fluctuations of the matter density along their line of sight. The cross-correlation of shear and galaxy cluster, often called cluster lensing, is related to cluster--matter cross-correlation. On large scales, the cluster-matter cross-correlation is linearly related to the matter two-point correlation function via the cluster bias.  Since the cluster lensing and cluster two-point correlations depend on the matter two-point correlation function with different powers of bias, they are complementary to each other: %
the combination of galaxy and cluster clustering with cluster weak lensing enables us to self-calibrate the clustering bias and measure the amplitude of matter fluctuations simultaneously.

This paper is organized as follows: 
In section~\ref{sec:simulation}, we detail the construction of cosmological survey simulations, including a brief summary of the creation of the mock catalogs, a description of the sample selection, and a comparison of galaxy cluster properties in different versions of the mock catalogs. In section~\ref{sec:selection}, we present a model describing the \redmapper{} selection bias, an additional large-scale bias that can be present if the \redmapper{} cluster finder preferentially selects clusters with properties that are correlated with the mass observables. We detail the model and covariance matrix calculation in section~\ref{sec:model}. In section~\ref{sec:measurement}, we describe the construction of the data vector in this analysis from galaxy and galaxy cluster catalogs. We summarize the analysis choice, including minimum scale cuts, estimation of samples' redshift distributions, the procedures to generate covariance matrices, and final model parameters %
in section~\ref{sec:analysis choice}.  In section~\ref{sec:robustness}, we test the robustness of the constraints on $\Omega_{\rm m}$ and $\sigma_8$ against various potential systematics. %
In sections~\ref{sec:simmain} and \ref{sec:cov}, we summarize the main results from analyzing simulated data sets, including an estimation of the theoretical systematic of our analysis pipeline, and a check of the validity of the theoretically derived covariance matrix employed in our analysis. Section~\ref{sec:conclusion} summarizes our findings.

 \begin{figure*}
\centering
\includegraphics[width=0.9\textwidth]{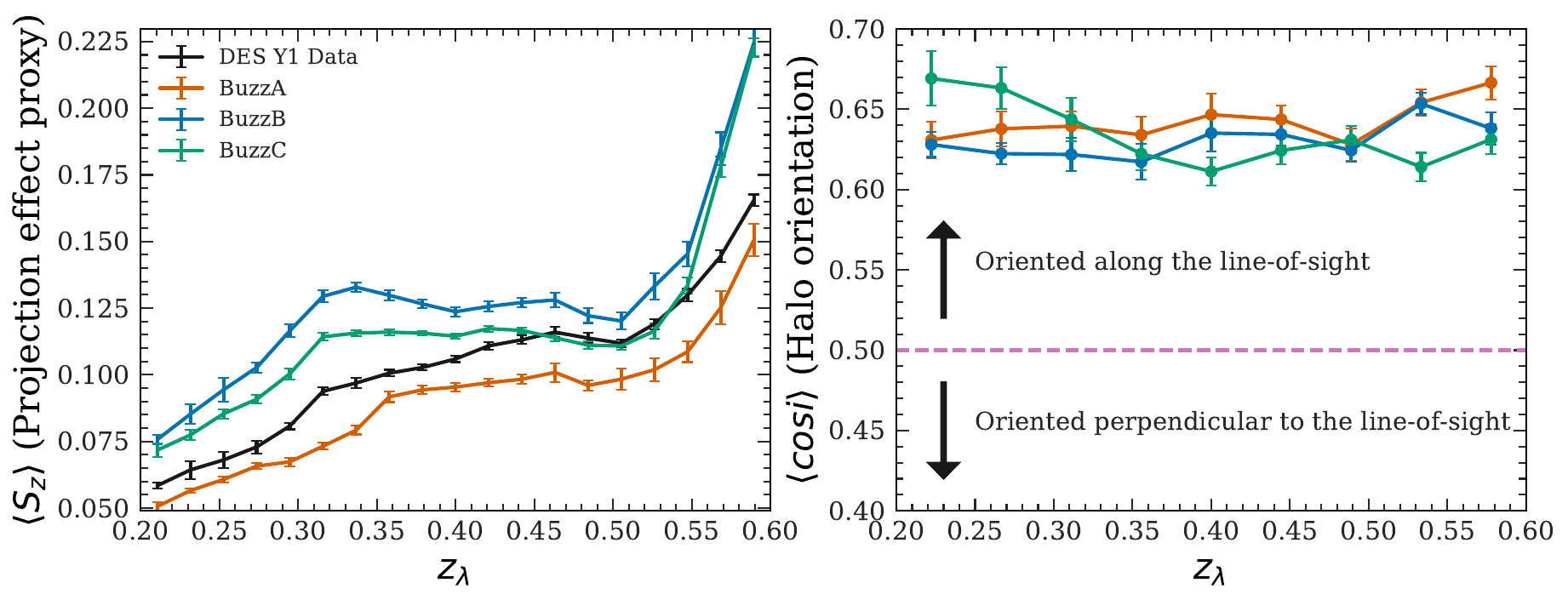}\hspace{-0.05\textwidth}
\caption{Projection effects and orientation biases in different versions of the simulations.
{\it Left-hand panel:} Mean $\rm{S}_z$ as a function of redshift. $\rm{S}_z$, defined in section~\ref{sec:simulation}, is expected to relate to the projection effects of \redmapper{} clusters.   %
The error bar is the error in the mean.
The {\it black} line represents the measurement in DES Y1 data \citepalias{DES_cluster_cosmology}. The {\it blue}, {\it green} and {\it orange} lines %
correspond to the measurement in each of the three different versions of \textsc{Buzzard}. The differences %
between these simulations are summarized in Table~\ref{tab:sims}. 
{\it Right-hand panel:} Mean \textbf{cosi} for \redmapper\ clusters, where \textbf{cosi} is the %
cosine of angle between the halo's major axis and the line of sight. The {\it blue}, {\it  green}, {\it orange} lines are %
the measurement in three different versions of \buzzard. %
Again, error bars represent the error on the mean.
In all three versions of \textsc{Buzzard}, \textbf{cosi} is greater than 0.5, indicating that the \redmapper{} cluster finder selects clusters that are preferentially oriented along the line of sight. The amount of orientation is consistent %
across simulations despite very different underlying galaxy--halo connections.
}
\label{fig:sigmaz}
\end{figure*}
\section{Simulations and Sample Selection}
\label{sec:simulation}
This analysis uses the \textsc{Buzzard} mock catalogs, which are described in detail in \cite{JoeBuzzard} and \cite{Addgal}. %
Here, we briefly summarize the key characteristics of the simulations and focus on the properties that are related to the performance of the cluster finder.

The creation of the \textsc{Buzzard} mock catalogs involves six steps. 
First, the N-body simulation is generated assuming a flat $\Lambda$CDM cosmology with $\Omega_{\rm m}=0.286$, $\sigma_8=0.82$, $\Omega_b=0.046$, $h=0.7$, and $n_s=0.96$. Second, the galaxies are populated into high resolution N-body simulations by the subhalo abundance matching model presented in \cite{2017ApJ...834...37L}, which matches brighter galaxies to halos with higher peak circular velocities while allowing for some scatter between the two.
Third, the model connecting galaxies' $r$-band absolute magnitudes ($M_r$) and local matter density is generated based on galaxies populated in the second step. In addition, a $M_r$--halo mass relation is fitted to central galaxies residing in resolved halos. Fourth, the low resolution N-body lightcone simulation is populated with galaxies in the following ways: central galaxies are populated on resolved halos according to the $r$-band magnitude--halo mass relation;  satellites and field galaxies are populated on dark matter particles according to the $r$-band magnitude--local matter density relation obtained from the previous step. Fifth, a spectroscopic sample of galaxies is used to populate a spectral energy distribution (SED) to each galaxy. This procedure is done by ranking galaxies in the spectroscopic data and galaxies in the simulation by their distances to the fifth nearest galaxy ($\Sigma_5$). In each magnitude bin, the SEDs of the spectroscopic galaxies are put on the simulated galaxies that have the same ranking. Sixth, we apply the DES survey depth and photometry uncertainties to each galaxy. We then run ray-tracing code {\sc CALCLENS}\footnote{\url{https://github.com/beckermr/calclens }} \citep{2013MNRAS.435..115B} to obtain the lensed magnitudes and galaxy shapes. The BPZ \citep{2000ApJ...536..571B} method is then run to obtain the photometric redshift of each galaxy (BPZ is the fiducial method used to estimate photometric redshifts of the source galaxies in \citetalias{DESY1KP, DES_cluster_cosmology}).

In this paper, we adopt the set of simulations created with this procedure as the baseline simulation (\buzzarda{}, version 1.9.2), which contains eleven realizations of the DES Y1 survey created from two sets of N-body simulations. In each realization, we run \redmagic{} and \redmapper{} on the galaxies in the same way as is done on the data. The end products are \redmagic{} galaxies, \redmapper{} clusters, and the shapes and photometric redshifts of all galaxies.

The galaxies in \buzzarda{} are found to have red-sequence colors with less scatter at fixed redshift than what is observed in the DES Y1 data (see Fig. 11. in \citealt{JoeBuzzard}). We expect that a less-scattered red sequence will lead
  to a more mild projection effect in \redmapper{} clusters. This is because a red sequence with less scatter helps \redmapper{} distinguish cluster galaxies
from foreground and background contamination. To verify this expectation,
we calculate the fraction of galaxies along the line of sight of a \redmapper{} cluster that would be counted as member galaxies of the given cluster as a function of their redshift separations. The width of this distribution $\rm{S}_z$, called $\sigma_z$ in \citetalias{DES_cluster_cosmology}, is expected to directly relate to the line-of-sight length scale within which \redmapper{} counts galaxies as cluster members.
This is also the quantity used to construct the projection effect model in \citetalias{DES_cluster_cosmology}. In this paper, we follow the same procedure as in \citetalias{DES_cluster_cosmology} and \cite{Matteoprojection}  to measure the $\rm{S}_z$ %
in the \textsc{Buzzard} mock catalogs. As shown in Fig.~\ref{fig:sigmaz}, the value of $\rm{S}_z$ %
for \redmapper{} clusters in \buzzarda{} is smaller than in the data, consistent with our expectation from the width of the red sequence in \buzzarda{}.

One of the goals of this paper is to test the robustness of the model developed in this paper against systematics introduced by the \redmapper{} cluster finder.
Therefore, we create a new simulation, \buzzardb{} (version 1.9.2+2), which increases the impact of projection effects relative to \buzzarda\ and even relative to DES data, thereby enabling a robust test of our systematics parameterization. The \buzzardb{} simulations are generated by adding Gaussian random noise to the color of red sequence galaxies in \redmapper{}. We then run \redmagic{}, \redmapper{}, and BPZ on the modified galaxy catalogs to obtain galaxy samples, cluster samples, and photometric redshift estimations of all galaxies. 

The differences between \buzzarda{} and \buzzardb{} test the robustness of the model against the amount of projection in the simulations. However, it doesn't test the robustness of the model against the assumptions of the galaxy--halo connection. %
To address this issue, we create \buzzardc{} (version 1.9.8) by making the following changes: first, the luminosity function used in the subhalo abundance matching is replaced by the luminosity function measured in the first three years of Dark Energy Survey data \citep[DES Y3][]{Y3Gold}. Second, we update the algorithm used to model color-dependent clustering. Instead of assigning SEDs of SDSS galaxies to our simulation by matching $\Sigma_5$ and $M_r$, we employ a conditional abundance matching scheme: galaxies at fixed $M_r$ in the SDSS data are ranked by their rest-frame $g$-$r$ color, and galaxies at fixed $M_r$ in our simulations are ranked by their distance to the nearest halo above a mass threshold, $M_h$. SDSS galaxies' SEDs are then assigned to simulated galaxies with the same rank as determined in the previous step, allowing for scatter in the relation between $g$-$r$ color and halo distance. The mass threshold, $M_h$ and the amount of scatter are tuned to fit measurements of $g$-$r$-dependent clustering in the SDSS Main Galaxy Sample \citep{2011ApJ...736...59Z}. We refer the reader to \cite{campaper} for further details. We note that \buzzardc{} contains one realization of the DES Y3 survey.

The differences between the simulations are summarized in Table~\ref{tab:sims}.  During our analysis, we found that one realization in both \buzzarda{} and \buzzardb{} behaves differently from the other realizations. 
In brief, we find that the \redmagic{} clustering in one
realization is anomalous, and in this realization a galaxy-clustering and galaxy--galaxy lensing analysis recovers a best-fit cosmology that is biased relative to the simulation by $3\sigma$.  A similar bias is observed for our cluster analysis.  Moreover, fixing galaxy biases to the measured values in the simulation, we find that in this one realization the \redmagic{} clustering returns cosmological constraints that are in $3 \sigma$ tension with the constraints from the galaxy--galaxy lensing. The cosmological constraints from galaxy clustering and galaxy--galaxy lensing in all other realizations recover the true cosmology within $1\sigma$. From these analyses, we conclude that the galaxy clustering in realization 3b is problematic, and therefore remove it from consideration for the rest of this paper. We caution that further analysis is needed to understand why this realization behaves differently from the others. Additional details are presented in appendix \ref{app:simulationrealization}.

\subsection{Sample selection}

We select two galaxy samples and one cluster sample from the simulations. 
The first galaxy sample %
is comprised of \redmagic{} galaxies, obtained by running the \redmagic{} algorithm \citep*{Redmagic} on the simulations with the same settings as %
the DES Y1 run. We then cut galaxies with redshift $z> 0.6$, the highest redshift of the \redmapper\ clusters.
We further split the galaxies into three bins using the
\redmagic{} photometric redshift ($z_{\rm{RMG}}$) estimate:
$0.15<z_{\rm{RMG}}<0.3$, $0.3<z_{\rm{RMG}}<0.45$, and $0.45<z_{\rm{RMG}}<0.6$. These redshift bins are consistent with the first three redshift bins of lens galaxies in \citetalias{DESY1KP}. Since we %
focus exclusively on galaxies with redshift less than $0.6$, we use the \redmagic{} high-density sample (luminosity, $L>0.5L_*$; number density, $n=1\times10^{-3}\ h^3\rm{Mpc}^{-3}$) %
for this analysis. 

The second galaxy sample consists of source galaxy samples.
Here, we do not run through the source galaxy selection procedure as described in \cite{2018MNRAS.481.1149Z}, which requires performing image simulations %
of the \textsc{Buzzard} mock catalogs. Instead, we use a procedure similar to %
that described in \cite{JoeBuzzard}, applying
size and magnitude
cuts to yield a similar source density as the DES Y1 data. The cuts we apply are:
\begin{enumerate}
    \item Mask all regions where the limiting magnitude and PSF size cannot be estimated, 
    \item $\sigma(m_{r,i,z})<0.25$,
    \item $\sqrt{r_{\rm{gal}}^2+r_{\rm{psf}}^2}>1.05r_{\rm{psf}}$, and
    \item $m_i<21.25+2.13z$,
\end{enumerate}
where $\sigma(m_{r,i,z})$ are magnitude errors in the $r,i,z$ bands, $r_{\rm{psf}}$ is the $i$-band PSF FWHM estimated from the data at the position of each galaxy, $r_{\rm{gal}}$ is the half light radius of the galaxy, and $z$ is the BPZ photo-$z$ of each galaxy. Note that these cuts are slightly different from the cuts in \cite{Niall, JoeBuzzard}. We find that these cuts reproduce better the galaxy number densities in the data.
We then use the BPZ photo-$z$ to split the samples into four redshift bins, defined as $0.2<z<0.43$, $0.43<z<0.63$, $0.63<z<0.9$, and $0.9<z<1.3$. 

The cluster samples are selected using the \redmapper{} \citep{Redmapper1} algorithm %
with the same settings as those described in \cite*{Tomclusterlensing}. We then split the \redmapper{} clusters into three redshift bins %
using the \redmapper{} photometric redshift: %
$0.2<z_{\rm{RMC}}<0.3$, $0.3<z_{\rm{RMC}}<0.45$, and $0.45<z_{\rm{RMC}}<0.6$. These redshift bins are chosen to maximize the redshift overlap between \redmapper{} clusters and \redmagic{} galaxies. Following \citetalias{DES_cluster_cosmology}, we split \redmapper{} clusters into four richness ($\lambda$) bins: $20<\lambda<30$, $30<\lambda<45$, $45<\lambda<60$, and $60<\lambda<\infty$.

\subsection{Comparison of properties of \redmapper{} clusters in different simulations}
\label{sec:redmappersystematic}

The properties of \redmagic{} galaxies and source samples in the \buzzard{} simulations are described extensively in \cite{Niall} and \cite{JoeBuzzard}. We refer the readers to those papers for details. Here we focus on the properties of \redmapper{} clusters. As pointed out in \citetalias{DES_cluster_cosmology}, two well-known systematics affecting the weak lensing signal of optically selected cluster are projection effects and orientation biases. The former is due to the imperfect separation of foreground and background galaxies  \citep{Tomomi}; the latter is due to the fact that \redmapper{} preferentially selects galaxy clusters when their major axes are aligned with the line of sight \citep{2014MNRAS.443.1713D, 2018MNRAS.477.2141O}. In this section, we compare these properties among the three versions of \textsc{Buzzard} and compare the simulation to the data where possible. In appendix \ref{app:datacomparison}, we show more comparisons of the simulation and the DES Y1 data. 

The amount of projection in the \redmapper{} catalog is related to the quantity $\rm{S}_z$ 
described in section~\ref{sec:simulation}. Fig.~\ref{fig:sigmaz} compares the mean $\rm{S}_z$ of our three sets of simulations %
to the measurement from the DES Y1 data \citepalias{DES_cluster_cosmology}. We find that the redshift dependence of $\rm{S}_z$ in the simulations is %
similar to that in the data.
Moreover, the $\rm{S}_z$ of the three simulations span the range of values of $\rm{S}_z$ in the data, suggesting that our simulations span an appropriately wide range of scenarios for the importance of projection effects in the data.

To test whether biases in halo orientation exist,  we measure \textbf{cosi}, the cosine of the angle between the halo's major axis and the line of sight. To avoid the uncertainty of associating clusters to halos (such as mis-centring), \redmapper{} is run by fixing the cluster center at the halo center. We then select clusters with richness greater than 20, the minimum richness cut for samples in this analysis, and measure their mean \textbf{cosi}. Fig.~\ref{fig:sigmaz} shows the comparison of mean
\textbf{cosi} for the three versions of \textsc{Buzzard}. %
We see all of the simulations predict that \redmapper{} clusters are preferentially aligned along the line of sight, consistent with similar findings in %
the literature \citep{2014MNRAS.443.1713D, 2018MNRAS.477.2141O}. We also note that despite having different galaxy--halo connection models, the three simulations in the analysis predict a very similar mean \textbf{cosi}. Because there is no measurement of this quantity, we can not assess whether our simulations have spanned the range that encompasses the data, though the consistency across multiple simulations suggests this is a robust prediction.

\begin{table*}

\begin{tabular}{lllll}
\hline
Simulation name   & \buzzarda{} & \buzzardb{} & \buzzardc{} \\ \hline
\textsc{Buzzard} version number & v1.9.2   & v1.9.2+2 & v1.9.8   \\
RedMaPPer mode                & Fullrun/Halorun & Fullrun/Halorun & Fullrun/Halorun \\ \hline
Footprint                    & DES Y1    & DES Y1    & DES Y3    \\
Survey depth                  & DES Y1    & DES Y1    & DES Y3    \\
Number of realizations         & 10      & 10     & 1        \\ \hline
\end{tabular}
\caption{Summary of the simulations adopted in this analysis. First, we employ three different galaxy--halo connection models: in \buzzarda{}, we employ our baseline model; in \buzzardb{}, we adjust the width of red sequence by adding scatters to red galaxies' luminosity; in \buzzardc{}, we adjust the color-dependent galaxy clustering and the width of red sequence. Second, in this analysis, we run \redmapper{} in two different modes.  In ``Fullrun,'' \redmapper{}  is run treating \textsc{Buzzard} galaxies as real data, and the run includes both cluster finding, center identification, and richness calculation. In ``Halorun,'' \redmapper{} is run fixing the cluster centers at the halo centers to avoid the ambiguity introduced when associating galaxy clusters to dark matter halos; richness is calculated at the halo centers. A comparison of these two different run modes of \redmapper{} is presented in appendix \ref{app:halorun_fullrun}. Halorun is used to develop the selection bias model. For the rest of the analysis in this paper, we use \redmapper{} catalogs in the Fullrun mode. 
}
\label{tab:sims}
\end{table*}

\section{Selection Effect of \redmapper{} clusters}
\label{sec:selection}
\begin{figure*}
\centering
\includegraphics[width=0.9\textwidth]{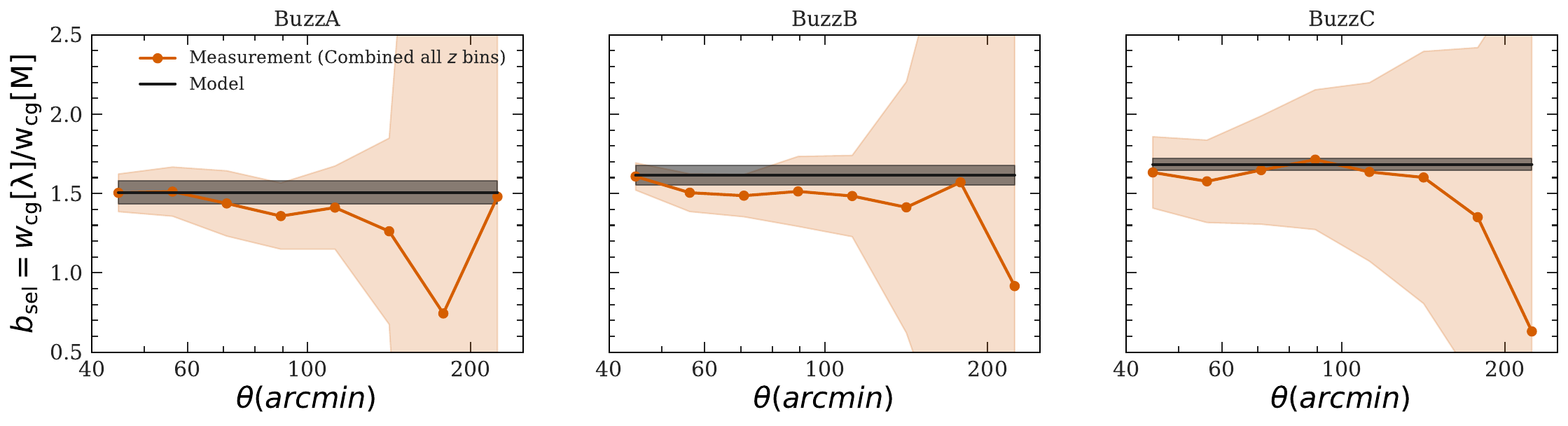}\hspace{-0.05\textwidth}
\caption{%
Ratios of the halo--galaxy cross-correlation functions between richness-selected halos, and halos re-weighted to match the mass and redshift distributions of the richness-selected halos.  The {\it orange} line denotes the mean of all realizations, with shaded areas showing the expected $1\sigma$ uncertainties error on the mean. Because there is only one realization of \buzzardc{}, the band corresponds to the theoretically expected uncertainty due to Poisson noise and sample variance.
Each panel shows the measurement of a version of the \textsc{Buzzard}mock catalogs summarized in Table~\ref{tab:sims}. The fact that the ratios deviate from 1 indicates the presence of a selection bias. %
The ratios are not scale dependent, allowing us to model %
them using a single parameter $b_{\rm{sel}}$. The black line corresponds to the best-fit theory model described in section~\ref{sec:selection}. The black shaded region 
corresponds to the $68$ per cent confidence interval estimated using the dispersion in the measurements of individual realizations within each family of simulations.}
\label{fig:selection_scale}
\end{figure*}

\begin{figure}
\centering
\includegraphics[width=0.5\textwidth]{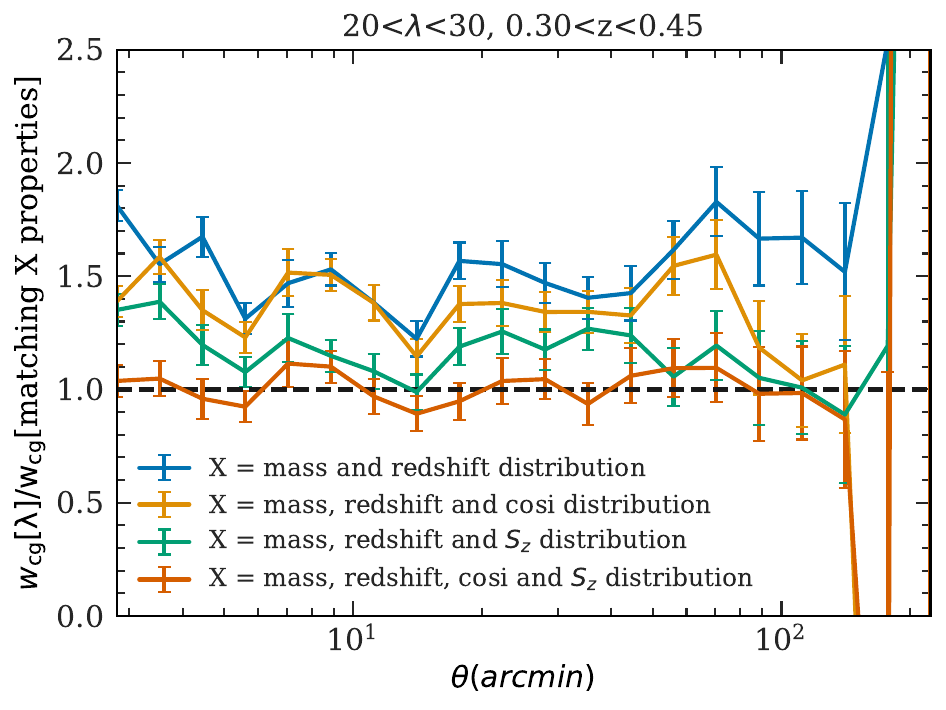}\hspace{-0.05\textwidth}
\caption{%
Ratios of the halo--galaxy cross-correlations between richness-selected halos and all halos with weights chosen to match X properties (shown in the legend) of the richness-selected halos. The error bars are 1-$\sigma$ errors estimated from 50 jackknife resamplings. %
Matching the halo mass and redshift of the \redmapper\ clusters is %
not sufficient to describe the clustering amplitude of the clusters.
Matching either the line-of-sight orientation or projection kernel $S_z$ alone improves the agreement between the reweighted halo sample and the \redmapper\ cluster selection, while matching both line-of-sight orientation and projection effects (as probed by $S_z$) is sufficient for reproducing the selection bias observed in the \redmapper\ clusters. Although this figure only shows one richness and redshift bin, the result holds for other redshift and richness bins.
 }
\label{fig:selection_appendix}
\end{figure}

As noted above, \redmapper\ entails important selection effects.  Because these selection effects also impact the cluster correlation function, the observable signal of the clusters depends not only on their mass, but also on the detailed quantitative impact of the \redmapper\ selection on the clustering statistics.
As we demonstrate below, over the scales used in this work, the selection effect manifests as an additional bias in the amplitude of the correlation functions. 
In the following, we refer to the selection effect introduced by the $\redmapper$ cluster finder as the \redmapper{} selection effect and the additional large scale bias of correlation functions due to this selection effect as selection bias. 
In this section, we measure the selection bias in the simulations. The goal is to develop a model to describe the \redmapper{} selection effect on cluster lensing, cluster--galaxy cross-correlations, and cluster auto-correlations. 

To better understand and quantify the \redmapper{} selection effect, we run \redmapper{} on sets of simulations with different galaxy--halo connection models. 
For the analysis of this section, \redmapper{} has been run fixing the cluster centers at the halo centers to avoid the ambiguity of associating galaxy clusters to dark matter halos. In Appendix \ref{app:halorun_fullrun}, we compare these cluster catalogs to those generated from the full \redmapper{} algorithm. There, we find that the differences between the two catalogs are small and do not impact our conclusions.

We start by examining the cluster--galaxy correlation function.  We compute the \redmapper{}--\redmagic{} cross-correlation functions in bins of richness and redshift (see section \ref{sec:measurement} for details). For each richness and redshift bin, we assign weights to all halos with $M_{200m} > 10^{13} h^{-1}M_{\odot}$, so that the weighted mass and redshift distribution %
of the halos is the same as that of the clusters in the bin. %
We then calculate the weighted halo--\redmagic{} cross-correlation functions and compare them to \redmapper{}--\redmagic{} cross-correlations. 
We refer to the ratio of these two correlation functions as the selection bias $b_{\rm sel}$.
Fig.~\ref{fig:selection_scale} shows the measured $b_{\rm{sel}}$ in the lowest richness bins, where we have the highest signal-to-noise ratio. It is %
clear that the %
selection bias deviates from 1, indicating that the samples are impacted by the selection effect. %
Moreover, Fig.~\ref{fig:selection_scale} also shows that $b_{\rm{sel}}$ is scale independent at the scales relevant to this project. We therefore model $b_{\rm{sel}}$ by a single scale-independent parameter%

In the simulation, we find that the measured $b_{\rm{sel}}$ appears to decrease from low to high richness, suggesting that $b_{\rm{sel}}$ might be mass dependent. We therefore model the selection bias as a power-law in mass, 
\begin{equation}
\label{eq:selection_bias}
    b_{\rm{sel}}(M) = b_{s0}(M/M_{\rm{piv}})^{b_{s1}},
\end{equation}
where $M_{\rm{piv}} = 5\times10^{14} h^{-1}M_\odot$,  $b_{s0}$ is the normalization, and $b_{s1}$ is the slope. We show the prediction of this model at the best-fit value obtained from the analysis of simulated catalogs compared to the measurements in Fig. \ref{fig:selection_scale}.

We assume that $b_{\rm{sel}}$ is redshift independent. This choice was made %
based on our analysis of \buzzarda{}, where no redshift evolution of $b_{\rm{sel}}$ is observed. 
In subsequent analysis of the \buzzardb{} and \buzzardc\ simulations we found that $3$ out of $11$ simulations exhibited redshift evolution at 2 to 3$\sigma$ significance, as determined from a direct fit to the galaxy and particle data.
While these realizations exhibit redshift evolution, the noise in the DES Y1 data set is sufficiently large that the bias on $\Omega_{\rm m}$ and $\sigma_8$ incurred from assuming no redshift evolution is small. In particular, in Fig. \ref{fig:countour_main}, we find that our posteriors are consistent with the input cosmology. We have also explicitly tested the impact of adding the redshift evolution in our posteriors through a reanalysis of the realization 4a of \buzzardb{}, the realization that exhibits the largest amount of redshift evolution among all realizations. 
Relative to the model that assumes redshift independent $b_{\rm{sel}}$, allowing for redshift evolution in $b_{\rm{sel}}$ shifts the posteriors toward the input cosmology. In particular, 
the smallest confidence contours containing the true $\sigma_8$ and $\Omega_{\rm{m}}$ parameters are the 81 per cent and 93 per cent confidence contours for the model with and without redshift evolution respectively.  The small difference in the contours demonstrates that these shifts are small relative to the statistical errors.
In the appendix \ref{app:bselz}, we detail the investigation of how the redshift-dependent $b_{\rm{sel}}$ affects the cosmological constraint.

We further investigate the connection between $b_{\rm{sel}}$ and the two known systematics in \redmapper{} clusters: projection effects and orientation biases, as described in section~\ref{sec:simulation}. 
Specifically, we reweight the halos so that in addition to matching the mass and redshift distributions of the \redmapper\ clusters, we also match the orientation and projection distributions of weighted halos and richness-selected halos %
as probed by \textbf{cosi} and $S_z$.
Fig.~\ref{fig:selection_appendix} shows that the halo--galaxy correlations of all halos with weights that match the mass, redshift, \textbf{cosi}, and $S_z$  distributions to the richness-selected halos is consistent with the halo--galaxy correlation of the richness-selected halos. This result indicates that the selection bias in \redmapper{}--\redmagic{} cross-correlations is due to projection effects and orientation biases, the two known dominant systematics in \redmapper{} samples. We expect that future work on quantifying these two systematics can put a tighter prior on the selection bias and hence tighten the cosmological constraints derived from the data. 

Here, although we measure the $b_{\rm{sel}}$ based on \redmapper{}--\redmagic{} cross-correlations, this selection bias is not limited to this part of the data vector. Given that galaxies are biased tracers of the dark matter density field, we expect the selection bias $b_{\rm{sel}}$ 
to apply for cluster--cluster and cluster--shear %
correlations %
as follows:
\begin{eqnarray}
\gamma_t [\rm{\lambda\ selected}] = b_{\rm{sel}}(M) \gamma_t [\rm{mass\ selected}]   \\
w_{cc} [\rm{\lambda\ selected}] = b_{\rm{sel}}^2(M) w_{cc} [\rm{mass\ selected}].
\end{eqnarray}
While we expect the above argument is valid on sufficiently large scales, %
we do not expect this simple model to hold at small scales.
For example, the \redmagic{} galaxies clustering signal may be correlated with the richness of \redmapper{} clusters at a fixed \redmapper{} mass. This correlation would introduce an additional \redmapper{} selection effect on \redmagic{}-\redmapper{} clustering, but not on cluster lensing and cluster clustering. We find that in the DES Y1 data \citepalias{DESY1KP, DES_cluster_cosmology}, the fraction of \redmagic{} galaxies in \redmapper{} clusters with richness above $\lambda = 20$ is $\approx 2$-$3$ per cent. Thus, we expect that %
any such \redmapper{} selection effect has %
negligible effects on the clustering of \redmagic{} galaxies scales greater than $8 h^{-1}{\rm{Mpc}}$, the minimum scale cut in this analysis. We also note that the above argument is only valid in the linear regime. Further analysis of the impact of selection bias on cluster lensing and cluster clustering beyond the linear regime needs to be done to extend this framework to small angular scales. %

\section{Model and Covariance Matrix}
\label{sec:model}
We assume the probability distribution $P(\bmath{D}|\bmath{p})$ of the observed data vector $\bmath{D}$ given the model parameters $\bmath{p}$ is Gaussian.  Therefore, the likelihood function takes the form,
\begin{equation}
\label{eq:likelihood}
    L(\bmath{p}|\bmath{D}) %
    \propto \rm{exp}
    \left(-\frac{1}{2} [\bmath{D}-\bmath{M(p)}]^T\textbfss{C}^{-1}[\bmath{D}-\bmath{M(p)}]\right),
\end{equation}
where $\bmath{M(p)}$ is the model prediction and $\textbfss{C}$ is the covariance matrix. In this section, we describe the construction of the model and the covariance matrix. 

\subsection{Model}
\label{sec:model_1}
The data vector of this analysis consists of the abundance of \redmapper{} clusters %
(N), as well as four distinct two-point correlations.  These are:%
(1) the auto-correlation of \redmagic{} galaxies %
$w_{\rm{gg}}(\theta)$; (2) the \redmapper{}-\redmagic{} cross-correlation $w_{\rm{cg}}(\theta)$; (3) the auto-correlation of \redmapper{} clusters  %
$w_{\rm{cc}}(\theta)$; and (4) the \redmapper{} %
cluster--shear cross-correlation $\gamma_{\rm{t, c}}(\theta)$. 

\subsubsection{Cluster abundance}

The \redmapper{} cluster abundance in a given richness ($\delta \lambda$) and redshift ($\delta z_i$) bin is given by

\begin{align}
\label{eq:counts}
     N^{i} = 
     &\int_{0}^{\infty}dz_{\rm{true}}\ 
     \frac{dV}{dz_{\rm{true}}} \langle \phi_i|z_{\rm true} \rangle \nonumber\\
     &\int_{\lambda\in\delta_\lambda} d\lambda
     \int_{0}^{\infty}dM\ 
     \mathbb{P}(\lambda | M, z_{\rm{true}}) \frac{dn}{dM}(M, z_{\rm{true}}), 
\end{align}
where
\begin{align}
     \langle \phi_i|z_{\rm true} \rangle  = 
     & \int_{z_{\rm{obs}}\in\delta z_i} dz_{\rm obs}\ p(z_{\rm obs}|z_{\rm true})\phi_i(z_{\rm obs})
\end{align}
and $\phi_i$ is the redshfit binning function for bin $i$, i.e. $\phi_i(z_{\rm obs})=1$ if $z_{\rm obs}$ is in bin $i$, and zero otherwise.
In the above expression, $z_{\rm{true}}$ denotes true redshifts of galaxy clusters, $M$ represents the cluster mass, $dV/dz_{\rm{true}}$  is the survey volume per unit redshift, and $dn/dM$ is the Tinker halo mass function \citep{Tinker10}. In Appendix~\ref{app:tinkeremulator}, 
we verify that replacing the Tinker mass function by an emulator \citep{HMFemulator} has a negligible impact on our results. 

In equation~\ref{eq:counts}, the redshift distribution $p(z_{\rm obs}|z_{\rm true})$ is averaged over all clusters,
\begin{align}
    p(z_{\rm obs}|z_{\rm true}) = A^{-1} \sum_a  {\cal N}(z_{\rm obs}|z_{{\rm RMC},a},\sigma^2_{z,a}),
\end{align}
where $A$ is a normalization constant; ${\cal N}(x|A,B)$ represents the Gaussian distribution with mean $A$ and variance $B$; $z_{\rm RMC}$ and $\sigma_z$ are reported photometric redshift estimation and uncertainty in the \redmapper{}; the sum is over all clusters $a$ in a thin redshift shell of width $\pm 0.005$ centered on $z_{\rm true}$. The survey volume per unit redshift is estimated by
\begin{equation}
    dV/dz_{\rm{true}} =\Omega_{\rm{mask}}(z)\frac{c}{H(z)}\chi^2(z),
\end{equation}
where $\chi$ represents the comoving distance, and $\Omega_{\rm{mask}}(z)$ is the effective survey area obtained from the \redmapper{} algorithm \citep{matteoSDSS}. We model the richness--mass relation ($\mathbb{P}(\lambda | M, z_{\rm{true}})$) as a log-normal model with scatter
\begin{equation}
   \sigma_{\rm{ln}\lambda}^2 = \sigma_{\rm{intrinsic}}^2+(e^{\langle \rm{ln}\lambda \rangle}-1)/e^{2\langle \rm{ln}\lambda \rangle}, 
\end{equation}
 and mean
 \begin{equation}
 \langle \ln(\lambda) |M \rangle = \rm{ln}\lambda_0 + A_{\lambda} \ln (M/M_{\rm{piv}}) + B_{\lambda} \ln ((1+z)/1.45).
 \end{equation}
 In section~\ref{sec:robustness}, we %
demonstrate that this model is sufficient to obtain unbiased cosmological constraints.

\subsubsection{Two-point clustering correlation functions}

A two-point correlation function can be related to the corresponding angular power spectrum via
\begin{equation}
    w^{i,j}_{\alpha \beta} (\theta) = \Sigma_\ell P_{\ell}(\rm{cos}(\theta))C^{i,j}_{\alpha, \beta}(\ell),
\end{equation}
where $\alpha$ and $\beta$ denote the two tracers being correlated (galaxy overdensity $g$ or cluster overdensity $c$), %
$i,j$ represents the tomographic bins of the two tracers, $\theta$ is the angular separation, and $P_{\ell}$ is the Legendre polynomial of order $\ell$.

In the linear regime, the angular power spectrum  $C^{i,j}_{\alpha, \beta}(\ell)$ of two density tracers at redshift bins $\delta z_i$ and $\delta z_j$ can be written as 
\begin{align}
\label{eq:linearcell}
    &C^{i,j}_{\alpha, \beta}(\ell) =\nonumber \\
    &\frac{2}{\pi}\int d z_1\  \int d z_2\  \int \frac{dk}{k}\ k^3\ \Delta^{\rm{NC},i}_\alpha(k, \delta z_i, z_1)\Delta^{\rm{NC},j}_\beta(k, \delta z_j, z_2) P_{\rm{lin}}(k, z_1,z_2),
\end{align}
where
\begin{eqnarray}
    &P_{\rm{lin}}(k, z_1,z_2)&=P_{\Phi}(k)T_{\delta}(z_1,k)T_{\delta}(z_2,k)\\
    &\Delta^{\rm{NC},i}_{\alpha}(k, \delta z_i, z) &= \Delta^{\rm{D},i}_{\alpha}(k, \delta z_i, z)+\Delta^{\rm{RSD},i}_{\alpha}(k, \delta z_i, z), \nonumber \\
    &\Delta^{\rm{D},i}_{\alpha}(k, \delta z_i, z) &=  q^{i}_{\alpha}(z)b^i_\alpha(z)j_{\ell}(k\chi(z)), \nonumber\\
    &\Delta^{\rm{RSD},i}_{\alpha}(k, \delta z_i, z) &= - q^{i}_{\alpha}(z)f(z)j_{\ell}''(k\chi(z)).
\end{eqnarray}
In the above expression, $b^i_\alpha(z)$ is the linear bias of the tracer $\alpha$ in redshift bin $i$, $q^{i}_{\alpha}(z)$ is the unit-normalized redshift distribution of the tracer $\alpha$, $k$ is the 3D wavenumber, $P_{\Phi}(k)$ is the power spectrum of the primordial curvature perturbations, $T_{\delta}$ is the matter overdensity transfer function, $f(z)=d\rm{ln}D/d\rm{ln}a$ is the scale independent growth rate, and $j''_\ell$ is the second order derivative of the spherical Bessel function.  The density tracer's number density $\Delta^{\rm{NC}}$ includes two contributions: a tracer's density contribution ($\Delta^{\rm{D}}$) and a linear contribution from redshift space distortions ($\Delta^{\rm{RSD}}$). We refer the reader to section 2.4.1 of \cite{CCL} for a more comprehensive description. In practice, we include the contribution of equal-time non-linear matter power spectra while evaluating the tracer's density contribution in equation \ref{eq:linearcell}. That is, we ignore the contribution of unequal-time  non-linear matter power spectra, which have been shown to be subdominant \citep{2019arXiv191111947F, 2019PhRvD.100b3543C}.

The unit-normalized redshift distribution of the \redmapper{} clusters ($q^{i}_{c}$) is calculated by 
\begin{equation}
\label{eq:redshift dist}
    q^{i}_{c}(z) = A^{-1}\langle \phi_i|z_{\rm true} \rangle\frac{dV}{dz}, 
\end{equation}
where $A$ is a normalization constant. Following \citetalias{DESY1KP}, we calculate the unit-normalized redshift distribution of \redmagic{} galaxy ($q^{i}_{g}$) by stacking $p(z|z_{\rm{RMG}})$, which is approximated by a Gaussian distribution with mean and $\sigma$ given by the redshift ($z_{\rm{RMG}}$) and photometric uncertainty reported by \redmagic{} algorithm. 

We treat the \redmagic{} galaxy bias $b_g(z)$ in each tomographic bin as a nuisance parameter, which is a constant in each redshift bin.  Unlike the galaxy bias, the bias of galaxy clusters is a predicted quantity in our model.%
We relate the bias to the mass of the galaxy clusters via measurement in N-body simulations \citep{Tinker10}. %
Here again replacing the Tinker bias by an emulator \citep{biasemu} has a negligible impact on our conclusions (see appendix \ref{app:tinkeremulator}).
As pointed out in section~\ref{sec:selection}, \redmapper{} clusters are subject to selection effects that manifest as an additional mass-dependent clustering bias. 
Thus, the net clustering bias of clusters in a given richness bin ($\delta \lambda$) at redshift $z$ is %
given by, 
\begin{align}
\label{eq:meanb1}
b^i_c(\delta \lambda, z) 
&= \langle b_Tb_{\rm{sel}} \rangle \nonumber\\
&= \frac{1}{n(\lambda,z)}\int d M\  b_T(M,z)b_{\rm{sel}}(M,z) \frac{dn}{dM}\int_{\lambda \in \delta \lambda} d \lambda\  \mathbb{P}(\lambda | M, z), 
\end{align}
where the normalization is given by
\begin{equation} 
{n(\lambda,z)=\int d M\  \frac{dn}{dM}\int_{\lambda \in \delta \lambda} d \lambda\  \mathbb{P}(\lambda | M, z)},
\end{equation}

$b_T(M,z)$ is the Tinker bias function, and $b_{\rm{sel}}(M,z)$ is the selection bias model defined in equation \ref{eq:selection_bias}. 

We evaluate equation \ref{eq:linearcell} using the fast generalized FFTLog\footnote{\url{https://github.com/xfangcosmo/FFTLog-and-beyond} } algorithm presented in \cite{2019arXiv191111947F}.

\subsubsection{Cluster lensing}
Cluster lensing is the measurement of the tangential shear of source galaxies around galaxy clusters. Here, we utilize the Limber approximation \citep{Limber} to convert the 3D power spectrum to the angular power spectrum. This analysis choice is justified in \cite{2019arXiv191111947F}, which shows that the galaxy--galaxy lensing model with Limber approximation is sufficiently precise to derive unbiased cosmological parameters from a Rubin Observatory %
LSST Y1-like survey. Given the large number density of galaxies relative to the number of galaxy clusters in a survey, as well as the steepness of the halo mass function relative to the bias--mass relation, the galaxy--galaxy lensing signal has a higher signal-to-noise than the cluster lensing signal at the same scale. Thus, we expect the Limber approximation to be sufficient for modeling the cluster lensing signal in this analysis.  Under the Limber approximation, the tangential shear of the background galaxies in redshift bin $j$ around the galaxy clusters in redshift bin $\delta z_i$ and richness bin $\delta \lambda$ at an angular separation $\theta$ can be written as \begin{align}
&\gamma_{\rm{t,c}}^{i,j}(\theta) =\nonumber\\ &\frac{3}{2}\Omega_{\rm m}\left(\frac{H_0}{c}\right)^2\int \frac{d\ell}{2\pi}\ \ell J_2(\ell\theta)\int dz\ \frac{g^j(z)q^{i}_{c}(z)}{a(z)\chi(z)}\avg{ P_{\rm{hm}}\left(k=\frac{\ell+1/2}{\chi(z)}, z\right)},
\end{align}
where $J_2$ is the second order Bessel function of the first kind, $a$ is the scale factor, $\chi$ is the comoving distance, $\avg{P_{\rm{hm}}}$ is the averaged cluster--matter power spectrum.
In the above expression $g^j(z)$ is the lensing efficiency for source galaxies in redshift bin $j$, computed as 
\begin{equation}
    g^j(z) = \int_z^\infty dz' q^{j}_{ s}(z')\frac{\chi(z')-\chi(z)}{\chi(z')},
\end{equation}
where $q^{j}_{ s}$ is the unit-normalized redshift distribution of source galaxies in redshift bin $j$, which is estimated using the BPZ photo-$z$ PDF estimates. 

Similar to equation \ref{eq:meanb1}, the averaged cluster--matter power spectrum $\avg{P_{\rm{hm}}}$ in redshift bin $i$ and richness bin $\delta \lambda$ can be written as 
\begin{equation}
    \avg{P_{\rm{hm}}(k,z)} =\frac{1}{n(\lambda,z)} \int d M\  P_{\rm{hm}}(k, M, z) \frac{dn}{dM}\ \int_{\lambda_{\rm{min}}}^{\lambda_{\rm{max}}} d \lambda\   \mathbb{P}(\lambda | M, z), 
\end{equation}
where $P_{\rm{hm}}(k, M, z)$ is the halo-matter power spectrum of halos with mass $M$ at redshift $z$.  

Following \cite{cosmolike2016}, the halo--matter power spectrum is modeled in the halo model framework \citep{2002PhR...372....1C}. In this model, $P_{\rm{hm}}(k, M, z)$ can be written as %
\begin{equation}
   P_{\rm{hm}}(k, M, z) = \frac{M}{\bar{\rho_m}}u(k,c,z)+b_T(M,z)b_{\rm{sel}}(M,z)P_{\rm{NL}}(k,z),
\end{equation}
where $\bar{\rho_m}$ is the mean matter density of the universe, $b_{\rm{sel}}$ is defined in section~\ref{sec:selection}, %
and $u(k,c,z)$ is the Fourier transform of the NFW profile with halo concentration $c$, for which we use the concentration--mass relation of \cite{2013ApJ...766...32B}. 

\subsection{Covariance Matrix}

 The Gaussian likelihood (equation \ref{eq:likelihood}) indicates that the covariance matrix is a key quantity that determines the error on the inferred cosmological parameters.  %
As summarized in \cite*{DESmethod}, the covariance matrix can be generated by three different methods: estimation from simulations, estimation from data, and analytical calculations. While the first two approaches require less theory assumptions, the covariance estimators are inherently noisy. The noise in covariance estimations leads to additional uncertainties to the inferred cosmological parameters estimated from the Gaussian likelihood \citep{2007A&A...464..399H, 2013PhRvD..88f3537D}. In this paper, we analytically compute the covariance matrix. This approach is motivated by the following arguments. First, unlike the estimation from simulations or data, there is no estimator noise in the theoretically derived covariance matrix, allowing the use of the Gaussian likelihood, instead of using a multivariate t-distribution \citep{2016MNRAS.456L.132S}. %
Second, as pointed out in \cite{HeidiCov}, the non-Gaussian terms in our covariance matrix are subdominant, and thus the corresponding uncertainties are not important.    

In this section, we summarize the analytic covariance matrix computation. The analytic covariance matrix can be separated into three components: angular two-point statistics x angular two-point statistics, angular two-point statistics x cluster abundance, and cluster abundance x cluster abundance. The covariance of two angular two-point functions $w_1, w_2 \in [w_{\rm{gg}}, w_{\rm{cg}}, w_{\rm{cc}}, \gamma_{\rm{t,c}} ]$ is related to the covariance of the angular power spectra by 
\begin{align}
    &\mathrm{Cov} \left(w_1^{i,j}(\theta),\, w_2^{k, m}(\theta')\right) = \nonumber\\ 
    &\int \frac{d\ell\ell}{2\pi} J_{n(w_1)}(\ell\theta)\int \frac{d\ell'\ell'}{2\pi} J_{n(w_2)}\left(\ell'\theta'\right)\left [\mathrm{Cov}(C^{i,j}_{w_1}(\ell),\,C^{k,m}_{w_2}(\ell'))\right],
\end{align}
where $n=0$ for $w_{\rm{gg}}, w_{\rm{cg}}, w_{\rm{cc}}$, and $n=2$ for $\gamma_{T,c}$. The term $\rm{Cov}\left(C^{i,j}_{w_1}(\ell),\,C^{k,m}_{w_2}(\ell')\right)$ is the covariance of angular power spectrum given by the sum of a Gaussian and a non-Gaussian covariance, including super-sample variance \citep{cosmolike2016}. The covariance of angular two-point functions and cluster abundance (N) can be related to the covariance of the angular power spectrum and cluster abundance via, 
\begin{align}
\mathrm{Cov}\left( w_1^{i,j}(\theta),\, N^i\right) = \int \frac{d\ell\ell}{2\pi} j_{n(w_1)}(\ell\theta)\left[\rm{Cov}(C^{i,j}_{w_1}(\ell),\, N^{i})\right],
\end{align}
where $\mathrm{Cov}\left(C^{i,j}_{w_1}(\ell),\, N^{i}\right)$ is the covariance of the angular power spectrum and cluster abundance. The cluster abundance cross cluster abundance terms are the sum of Poisson shot noise terms and super-sample variance terms. We refer the reader to \citet{cosmolike2016} for more details.

\section{Results}
\label{sec:result}
\subsection{Measurement}
\label{sec:measurement}
We measure the two-point correlation functions --- galaxy clustering, galaxy--cluster cross-correlations, cluster clustering --- using the Landy--Szalay estimator \citep{1993ApJ...412...64L}, 
\begin{equation}
    \hat{w}(\theta) = \frac{DD-2DR+RR}{RR}, 
\end{equation}
where $DD$ is the number of pairs of tracers (galaxies or galaxy clusters) with angular separation $\theta$, $RR$ is similarly defined for a catalog of points whose positions are randomly distributed within the survey volume (random points), and $DR$ is the number of cross pairs between tracers and random points. 
The correlation functions are calculated in 20 logarithmic angular bins between 2.5 to 250 arcmin to match the analysis in \citetalias{DESY1KP}. The pair counting is done by {\sc Corrfunc}\footnote{\url{https://github.com/manodeep/Corrfunc}}  \citep{2020MNRAS.491.3022S}.

The cluster lensing tangential shear signal $\gamma_{\rm{t,c}}^{i,j}(\theta)$ is measured by averaging the tangential shear ($e_{\rm T}$) of source galaxies over all cluster--source galaxy pairs with an angular separation $\theta$. The  $\gamma_{\rm{t,c}}^{i,j}(\theta)$ estimator is written as
\begin{equation}
\label{eq:lensing estimator}
    \hat{\gamma}_{\rm{t,c}}^{i,j}(\theta) = \frac{\sum_{\alpha \in \rm{DS}^{i,j}(\theta) }e^\alpha_{\rm{T}}}{\rm{DS}^{i,j}(\theta)}-\frac{\sum_{\alpha\in\rm{RS}^{i,j}(\theta) }e^{\alpha}_{\rm{T}}}{\rm{RS}^{i,j}(\theta)},
\end{equation}
where $\rm{DS}^{i,j}\left(\theta\right)$ is the number of cluster--source galaxy pairs of clusters at redhsift bins $i$ and source galaxies at redshift bins $j$ that are separated by an angular separation $\theta$, and $\rm{RS}^{i,j}\left(\theta\right)$ is similarly defined as $\rm{DS}^{i,j}\left(\theta\right)$ but on random--source galaxy pairs. $e^\alpha_{\rm{T}}$ is the tangential shear of the source galaxies in cluster--source galaxy pair $\alpha$. %

This estimator is biased due to photometric redshift uncertainties. Due to uncertainties 
in the redshift estimations, some of the source galaxies are members of galaxy clusters. These galaxies are not lensed by galaxy clusters, thus diluting the lensing signal. %
\cite{2004AJ....127.2544S} point out that this dilution effect can be measured by the following estimator: 
\begin{equation}
    B^{i,j}(\theta) = \frac{N^i_r}{N^i_c}\frac{DS^{i,j}(\theta)}{RS^{i,j}(\theta)}, 
\end{equation}
where $N_r$ is the number of random points, $N_c$ is the number of galaxy clusters and  $RS^{i,j}(\theta)$ is the number of random points--source galaxy pairs with angular separation $\theta$. The $B^{i,j}
(\theta)$ is usually called boost factor in the literature, and $1/B^{i,j}(\theta)$ is the amount of dilution due to photometric uncertainties. Since the boost factor is measured in the data, we apply this correction directly on the estimator. Using this correction, our cluster lensing estimator is written as
\begin{eqnarray}
    \hat{\gamma}_{\rm{t,c}}^{i,j}(\theta) &=& \frac{\sum_{\alpha \in \rm{DS}^{i,j}(\theta) }e^\alpha_{\rm{T}}}{\rm{DS}^{i,j}(\theta)} \frac{N^i_r}{N^i_c}\frac{DS^{i,j}(\theta)}{RS^{i,j}(\theta)} -  \frac{\sum_{\alpha\in\rm{RS}^{i,j}(\theta) }e^{\alpha}_{\rm{T}}}{\rm{RS}^{i,j}(\theta)}. \nonumber \\
    &=& \frac{N^i_r}{N^i_c} \frac{\sum_{\alpha \in \rm{DS}^{i,j}(\theta) }e^\alpha_{\rm{T}}}{\rm{RS}^{i,j}(\theta)}-  \frac{\sum_{\alpha\in\rm{RS}^{i,j}(\theta) }e^{\alpha}_{\rm{T}}}{\rm{RS}^{i,j}(\theta)}.
\end{eqnarray}
We note that since there is no lensing effect around random points, this term is not subject to the dilution due to the photometric redshift uncertainties. Thus, we do not apply the boost factor correction on the second term of equation~\ref{eq:lensing estimator}. 

The cluster lensing ($\gamma_{\rm{t,c}}^{i,j}(\theta)$)  are calculated in 20 logarithmic angular bins between 2.5 to 250 arcmins. The calculation is done by {\sc Treecorr}\footnote{\url{https://github.com/rmjarvis/TreeCorr}} \citep{2004MNRAS.352..338J}.

\subsection{Analysis choices}
\label{sec:analysis choice}
We summarize our analysis choices below. We expect these analysis choices to be carried through to the analysis of real data.

\begin{enumerate}
    \item \textbf{Minimum angular scale cuts.}
    For $w_{\rm{gg}}(\theta)$, $w_{\rm{cg}}(\theta)$, and $\gamma_{\rm{t,c}}(\theta)$, we adopt a minimum scale cut corresponding to $8h^{-1}\rm{Mpc}$ at the mean redshift of each lens redshift bins. In section~\ref{sec:robustness}, we justify this scale cut by verifying that the cosmological posteriors derived from our analysis are robust to a variety of systematics when adopting this cut. For $w_{\rm{cc}}(\theta)$, we adopt a minimum scale cut corresponding to $16h^{-1}\rm{Mpc}$ at the mean redshift of each cluster redshift bin. The scale cut is chosen such that the $\chi^2$ of the best-fit model is consistent with the $\chi^2$ between data vectors measured in different realizations of the same simulation scheme. In this way, we obtain the minimum scale where the best-fit model provides a good description of the $w_{\rm{cc}}(\theta)$ without relying on the exact value of the theory covariance matrix. We further confirm that the distribution of $\chi^2$ values is consistent with the expected $\chi^2$ distribution based on the number of degrees of freedom. Finally, we apply an additional scale cut on $w_{\rm{cc}}(\theta)$ to avoid biases and large fluctuations to the correlation function measurement from to sparseness issues when there are only few cluster pairs in the angular bins. Thus, we cut out the angular bins when the expected number of pairs are less than one hundred.  We find that this additional scale cut largely improves the $\chi^2$ of the best-fit model. 
    
    \item \textbf{Redshift Distributions.} The redshift distributions of the lens samples ($\hat{P}_{\delta z}(z)$) are calculated based on photometric redshifts estimated by \redmapper{} and \redmagic{}. The redshift distributions of the source galaxies are estimated from photometric redshifts estimated by BPZ \citep{2000ApJ...536..571B}. Following \citetalias{DESY1KP}, we introduce two sets of nuisance parameters to account for systematics in photometric redshift estimations. The systematics are modeled through shift parameters $\Delta^i_{z,\alpha}$, so that 
    \begin{equation}
        q^i_{\alpha}(z) = \hat{q}^i_{\alpha}(z-\Delta^i_{z,\alpha}), \ \alpha \in \{g,s\},
    \end{equation}
    where g denotes \redmagic{} galaxies, s denotes source galaxies, and  $\hat{q}^i_{\alpha}(z-\Delta^i_{z,\alpha})$ denotes the estimated redshift distributions based on photometric redshifts. Note that we do not account for the redshift systematic of \redmapper{} clusters, since \citetalias{DES_cluster_cosmology} demonstrate that this systematic is subdominant.  These shift parameters $\Delta_{z, \alpha}$ are marginalized over using Gaussian priors of width [0.008,0.007,0.007] for \redmagic{} galaxies and [0.016,0.013,0.011,0.022] for source galaxies. The mean of the Gaussian prior is estimated by comparing the true redshift distribution in simulations and the photometric redshift estimations. This is clearly not possible in a real data analysis. In the analysis of real data, the mean of the Gaussian prior is estimated using cross-correlations of galaxy samples with spectroscopic samples \citep*{2018MNRAS.478..592H, 2018MNRAS.481.2427C}. Because we focus on cluster-related systematics in this paper, we do not repeat this process in the simulation.
   \item \textbf{Matter power spectrum.}
    We evaluate the non-linear matter power spectrum  using the  \cite{1998ApJ...496..605E} approximation for the transfer function and the revised \textsc{HALOFIT} fitting formula of  \cite{Takahashi2012} for the non-linear evolution. To validate this model, we compare the theory data vector generated at the true cosmology to that generated from \textsc{CLASS} \citep{CLASS} and \textsc{HALOFIT}. We find that the $\Delta \chi^2$ between the two data vectors is $1.45$.  Thus, we conclude that this theory approximation does not affect the conclusion of this paper. 
\item \textbf{Theory covariance matrix.} The covariance matrix is calculated assuming a fixed set of cosmological and nuisance parameters. %
\cite{2013A&A...551A..88C} shows that %
when approximating the true data likelihood with a Gaussian likelihood, the parameter posteriors better match the true uncertainty in the measurement when the cosmological dependence of the covariance matrix is ignored.  In particular, allowing the covariance matrix to vary with cosmology results in over-optimistic constraints.
 In this analysis, we fix the cosmological parameters for the covariance matrix at the true cosmology in the \textsc{Buzzard} mock catalogs. This is clearly not possible in an analysis of real data. However, \citetalias{DESY1KP} shows that there is negligible change in the parameter constraints in the 3$\times$2pt analysis while using two different cosmologies to calculate the covariance matrix. Because our data vectors are more shot-noise dominated than the 3$\times$2pt data vectors, we expect our conclusions to be insensitive to this analysis choice. 

Unlike a 3x2pt analysis, however, our observable also depends on the richness--mass relation and the selection bias, for which we do not have good {\it a priori} estimates.
We use an iterative approach to obtain the richness--mass relation %
and selection bias parameters used to compute the covariance matrix. 
We start by setting $b_{s0}=1$ and $b_{s1}=0$.
The fiducial richness--mass relation is obtained from fitting the cluster abundance data only
assuming only shot noise, and adopting the true cosmology of the simulations. This selection bias and richness--mass relation are then used to generate a covariance matrix, which is adopted while fitting cluster abundance and cluster clustering simultaneously at the true cosmology of the simulation.   %
We update the richness--mass relation and selection bias parameters with the best-fit parameters, and re-fit the cluster abundance and cluster clustering data.  Thus, we obtain a new estimate of the richness--mass relation and selection bias parameters, enabling us to construct a new covariance matrix and perform a new fit. We keep iterating until convergence.%
We find that the $\chi^2$ converges after three iterations.  When calculating the covariance matrix for the real data, we hold the cosmology fixed to the best-fit cosmology in the  \cite{DESY1KP} paper. We %
show in appendix \ref{app:covariance} that 
our recovered cosmological constraints are largely insensitive to the differences between the various covariance matrices in this iterative procedure. We note that the iteration processes are performed for each DES realization independently.

\item \textbf{Free parameters.} For the nuisance parameters, we marginalize over three linear galaxy bias parameters, one for each redshift bin.  Likewise, our model contains three shift parameters characterizing the redshift bias of \redmagic\ galaxies, and
four shift parameters for the redshift bias of source galaxies.
Finally, our model has four richness--mass relation parameters, and two selection bias parameters. %
Here, we do not consider intrinsic alignment effect for the following reasons. First, the effect is expected to be small for cluster lensing \citep{2015A&A...575A..48S}. Second, while modeling cluster lensing, we exclude bins where the maximum redshift of galaxy clusters is larger than the mean redshift of source galaxies. 
We marginalize over the same set of cosmological parameters as in \citetalias{DESY1KP} (and using the same priors), except for the sum of neutrino mass and $A_s$. Because the sum of neutrino mass is zero in the simulation, the wide prior adopted in the \citetalias{DESY1KP} analysis would lead us to recovered biased cosmological parameter estimates in the simulation \citep*{Niall, DESmethod}. Further, since we adopt \cite{1998ApJ...496..605E} approximation for the transfer function, we %
sample $\sigma_8$ instead of $A_s$. Note in particular that this means that our priors are flat in $\sigma_8$ as opposed to flat in $A_s$. We have verified that this analysis choice has negligible impact on the constraints on $\Omega_{\rm m}$ and $\sigma_8$.

The parameters and priors are summarized in Table~\ref{tab:paramsumm}.
    
\end{enumerate}
\begin{table}
    \centering
    \footnotesize
    \caption{Parameters and priors considered in this analysis. Flat represents the flat prior in the given range and $\rm{Gauss}(\sigma)$ denotes the Gaussian prior with width $\sigma$. The means of the Gaussian priors are determined by comparing true redshifts and photometric redshifts of galaxies, thus varying between different versions of simulations.}
    \label{tab:paramsumm}
   \begin{tabular}{|| lccc ||}
    \hline 
	Parameter	 & Prior\\
	\hline 
	&Cosmology& \\
	$\Omegam$ & Flat (0.1,0.9) \\
	$\sigma_8$ & Flat (0.5,1.2) \\
	$n_s$ & Flat (0.87, 1.07)\\
	$\Omega_b$ &  Flat (0.03, 0.07)\\
	$h$ &  Flat (0.55, 0.91) \\
	\hline 
	&Galaxy Bias& \\
	$b_1^1$ &  Flat (0.8, 3.0)\\ 
	$b_1^2$ &  Flat (0.8, 3.0)\\ 
	$b_1^3$ &  Flat (0.8, 3.0)\\ 
	\hline
	&\redmagic{} photo-$z$&\\
	$\Delta_{z,g}^1$ & $\rm{Gauss}(\sigma=0.008)$ \\
	$\Delta_{z,g}^2$ & $\rm{Gauss}(\sigma=0.007)$ \\
	$\Delta_{z,g}^3$ & $\rm{Gauss}(\sigma=0.007)$ \\
	\hline
	&Source galaxy photo-$z$& \\
	$\Delta_{z,s}^1$& $\rm{Gauss}(\sigma=0.018)$ \\
	$\Delta_{z,s}^2$ &$\rm{Gauss}(\sigma=0.013)$ \\
	$\Delta_{z,s}^3$& $\rm{Gauss}(\sigma=0.011)$ \\
	$\Delta_{z,s}^4$& $\rm{Gauss}(\sigma=0.022)$ \\
	\hline
    &\redmapper{} richness--mass relation \\
    $\rm{ln}\lambda_{0}$ &  Flat (2.0,5.0)\\
    $A_{\rm{ln}\lambda}$ &  Flat (0.1,1.5)\\
    $B_{\rm{ln}\lambda}$ & Flat(-5.0, 5.0)\\
    $\sigma_{\rm{intrinsic}}$ &  Flat(0.1, 1.0)\\
    \hline 
    &\redmapper{} selection effect \\
    $b_{s0}$ &  Flat (1.0,2.0)\\
    $b_{s1}$ & Flat(-1.0,1.0)\\
   \hline \vspace{-3mm}\\   
    \end{tabular}
\end{table}

\subsection{Robustness}
\label{sec:robustness}

We quantify the impact of potential systematics that %
cannot fully 
be tested by the simulations. %
Specifically, we consider the following systematics: small-scale lensing systematics, the functional form of the richness--mass relation, and beyond linear bias expansions. In all cases, our robustness tests follows the same procedure:
\begin{enumerate}
    \item Generate a noiseless data vector that includes the systematic effect being tested. 
    \item Analyze this data vector with the baseline model. 
    \item Measure the bias in $\Omega_{\rm m}$ and $\sigma_8$ due to the unaccounted systematics.
\end{enumerate}
All analyses are done using the analytical covariance matrix
generated using the best-fit cosmology of \citetalias{DESY1KP} and the richness--mass relation parameters in \cite{Lindsy}. 

\begin{figure}
\centering
\includegraphics[width=0.5\textwidth]{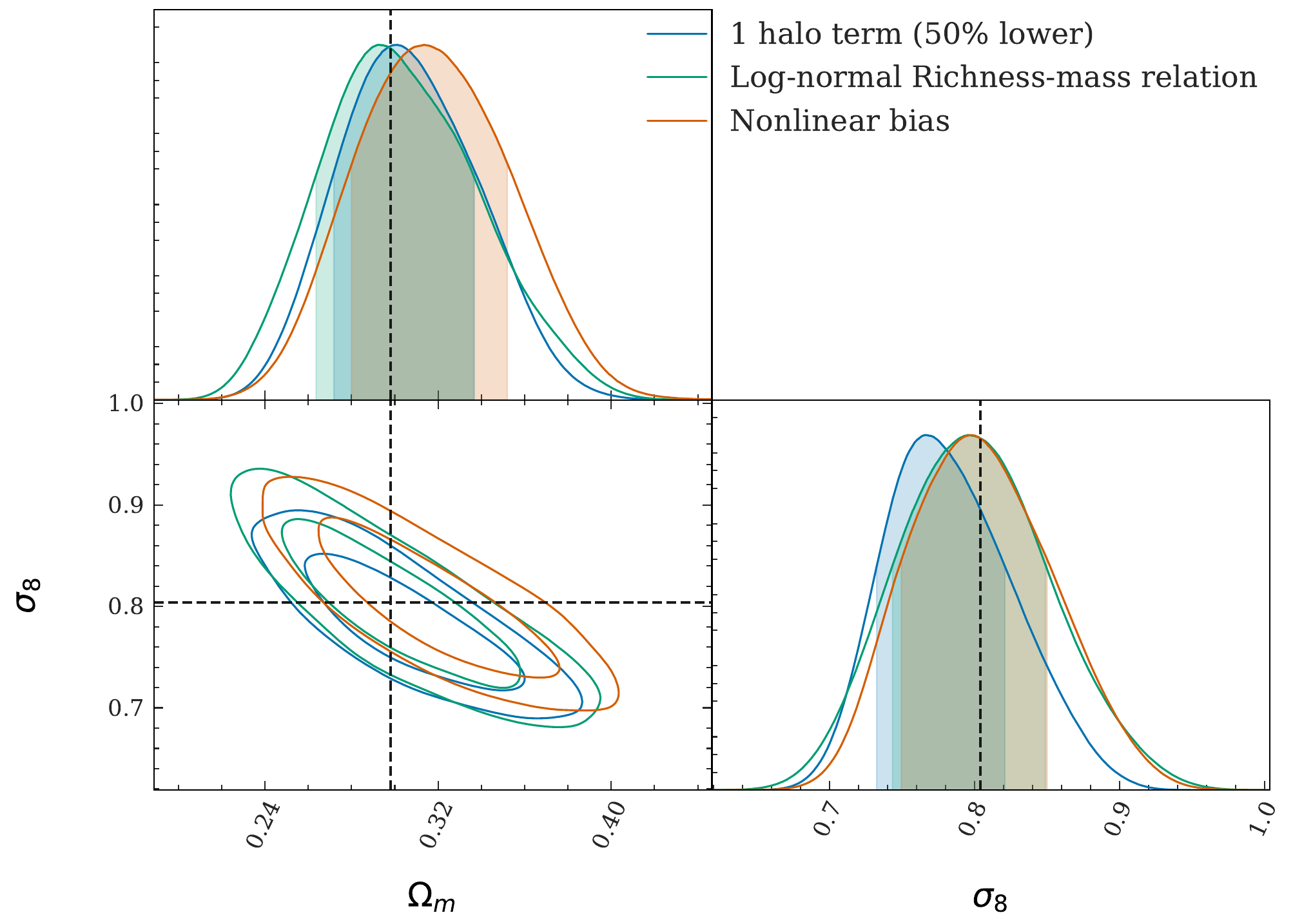}\hspace{-0.05\textwidth}
\caption{Biases in the parameters $\Omega_{\rm m}$ and $\sigma_8$ due to different systematic uncertainties. The contours show the $68$ per cent and $95$ per cent confidence levels, corresponding to the $1\sigma$ and $2\sigma$ expected statistical uncertainties of a DES-Y1 like survey. The dashed line indicates the cosmology of the input data vectors. We test three possible systematics. %
The {\it blue} line shows the parameter biases due to the possible $50$ per cent underestimation of the cluster lensing one-halo term, a potential systematic suggested in \citetalias{DES_cluster_cosmology}. The {\it orange} line shows the parameter biases due to the unaccounted-for non-linear galaxy and galaxy cluster biases. The {\it green} line shows the parameter biases due to the unaccounted-for complicated richness--mass relation. The figure shows that our result is unbiased due to each of these three systematics. %
}
\label{fig:systematics}
\end{figure}

Fig.~\ref{fig:systematics} quantifies the cosmological parameter bias due to these three systematic effects that are not fully captured by the \textsc{Buzzard} mock catalogs. The first systematic is the sensitivity to the anomalously low lensing signal at small scales for low richness clusters described in \citetalias{DES_cluster_cosmology}.  A similar low lensing signal has been detected in analyses of SDSS spectroscopic galaxies \citep{Lensingislow}.
Importantly, \citetalias{DES_cluster_cosmology} %
points out that the \textsc{Buzzard} mock catalogs do not %
exhibit a similar feature.  We test the sensitivity of our analysis to this systematic by reducing the amplitude of the one-halo term of our lowest richness bins by 50 per cent.

The second systematic is the functional form of the richness--mass relation. The richness--mass relation depends on both the %
galaxy--halo connection as well as the performance of the cluster finder \citep{Matteoprojection}. %
If the functional form of the richness--mass relation in the simulations does not reproduce that of the data, then our simulation tests may leave us blind to this possible source of systematic uncertainty.
For example, \cite{JoeBuzzard} finds a deficit in richness at fixed halo mass when comparing \buzzard{} to DES Y1 data \citep*{Tomclusterlensing}, leading to a factor of 2 %
fewer $\lambda>20$ galaxy clusters in \buzzard{} than in the data.  
Thus, while the log-normal richness--mass relation is sufficient to describe the richness--mass relation in the \textsc{Buzzard} mock catalogs, it does not guarantee %
that it will adequately describe the data.
To check the robustness of our cosmological constraints against this systematic, we generate the input data vector using the richness--mass relation %
in \citetalias{DES_cluster_cosmology}, %
which we then analyze with our baseline log-normal richness--mass relation model. 

The third systematic is the possible presence of non-linear galaxy and galaxy cluster biases. This systematic is partially tested by the simulations; however, the size of the effect depends on the mean mass of the galaxy cluster, thus depending on the exact value of the richness--mass relation. Because we do not expect the simulation to perfectly reproduce the richness--mass relation of the real data, we check the robustness of this model using a theoretical calculation. The input data vector is generated including the next-to-leading order contribution from quadratic bias $b_2$, tidal bias $b_s$, and the third-order non-local bias $b_{\rm{3nl}}$ \citep{2009JCAP...08..020M, 2012PhRvD..86h3540B}. The nonlinear contributions are evaluated using the \textsc{FAST-PT} code \citep{2016JCAP...09..015M} with $b_2$ determined by $b_2-b_1$ relation measured in N-body simulations \citep{2016JCAP...02..018L}. The $b_s$ and $b_{\rm{3nl}}$ are determined by their relation to $b_1$ derived from  the equivalence of Lagrangian and Eulerian perturbation theory \citep{2014PhRvD..90l3522S}.
This data vector is then analyzed by the baseline linear bias model. 

Fig.~\ref{fig:systematics} shows that none of the above %
systematics %
bias our posteriors by more than 0.5$\sigma$. We therefore conclude that our model is sufficiently flexible to enable us to derive robust cosmological constraints at the precision achievable by DES Y1-like surveys. More detailed modeling may be required for future, more constraining, analyses. 

In this analysis, we do not consider \redmapper{} mis-centring as a potential systematic for two reasons. %
First, the scale of mis-centring of \redmapper{} clusters is $\approx 0.2$ $h^{-1}\rm{Mpc}$ \citep{miscentering}, %
much smaller than the smallest scales included in our analysis. %
Second, any additional scatter in richness estimates due to miscentering effects can be absorbed by the richness--mass scaling relation parameters (section~\ref{sec:model_1}). 

\begin{table*}
\begin{tabular}{lllll}
\hline
Simulations   & \buzzarda{} &\buzzardb{}  &\buzzardc{}  \\ \hline
$\Delta \Omega_{\rm m} $ & $0.037 \pm 0.016$ &
$0.043 \pm 0.019$&
$0.080 \pm 0.037$\\
$\Delta \sigma_8 $ & $-0.056 \pm 0.022$ &
$-0.061 \pm 0.028$ &
$-0.055 \pm 0.048 $\\
\hline
 minimal confidence interval encompasses $\Delta \theta=0$& 0.969& 0.930& 0.992\\
\hline
$P(\rm{sys}<\sigma_{\rm{Y1}})$ & 0.65 &0.57 &0.19  \\ 
$P(\rm{sys}<2\sigma_{\rm{Y1}})$  & 0.97 & 0.93 &0.55  \\
 \hline
\end{tabular}
\caption{A summary of constraints on the size of parameter biases in $\Omega_{\rm m}-\sigma_8$ parameter spaces inferred from different versions of the \textsc{Buzzard} mock catalogs. The first two rows show the bias in $\Omega_{\rm m}$ and $\sigma_8$ with $68$ per cent uncertainties. The third row shows the minimal confidence interval containing the null hypothesis ($\Delta \theta=0$).  The forth and fifth rows show the probability that the systematic is smaller than the expected DES-Y1 $68$ per cent and $95$ per cent uncertainties. }
\label{tab:mainsum}
\end{table*}

\begin{figure*}
\includegraphics[width=0.4\textwidth]{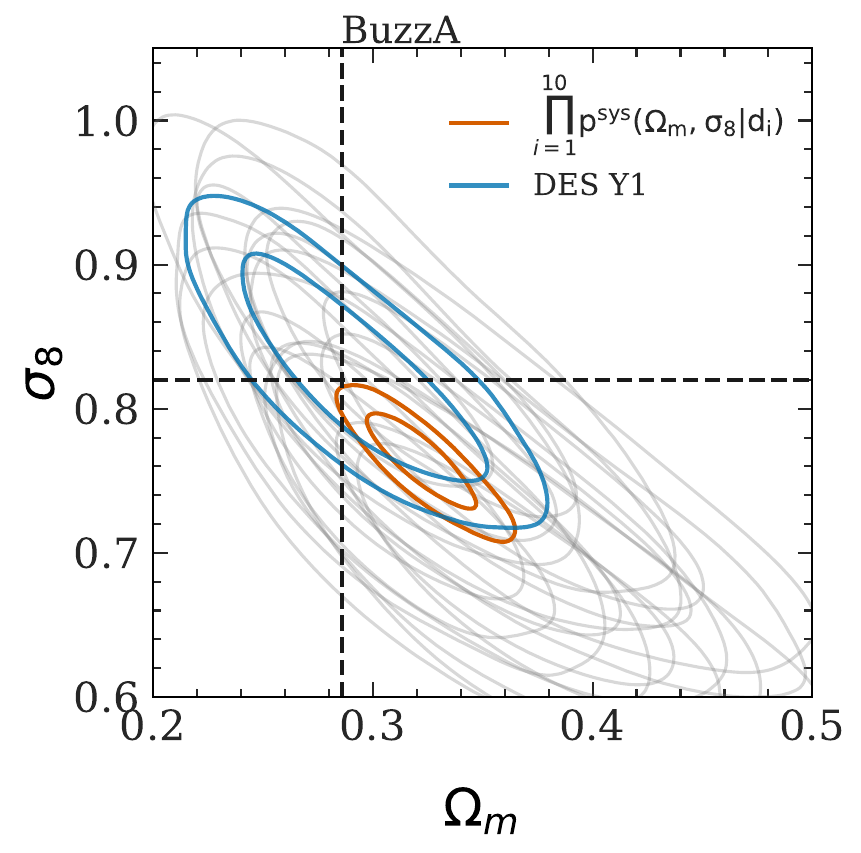}
\includegraphics[width=0.4\textwidth]{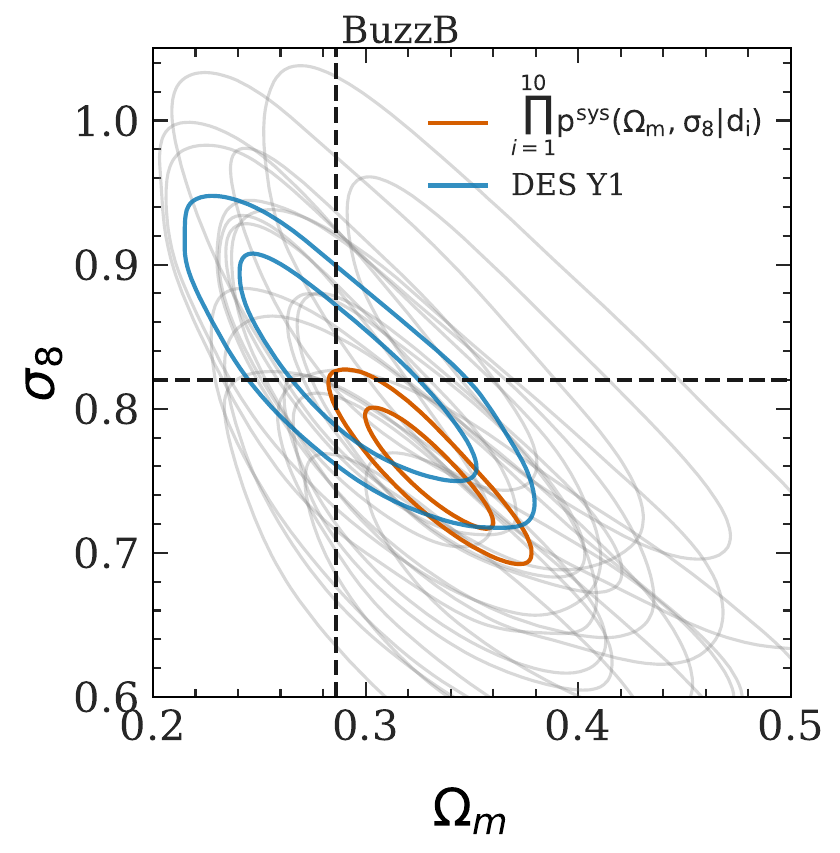}
\includegraphics[width=0.4\textwidth]{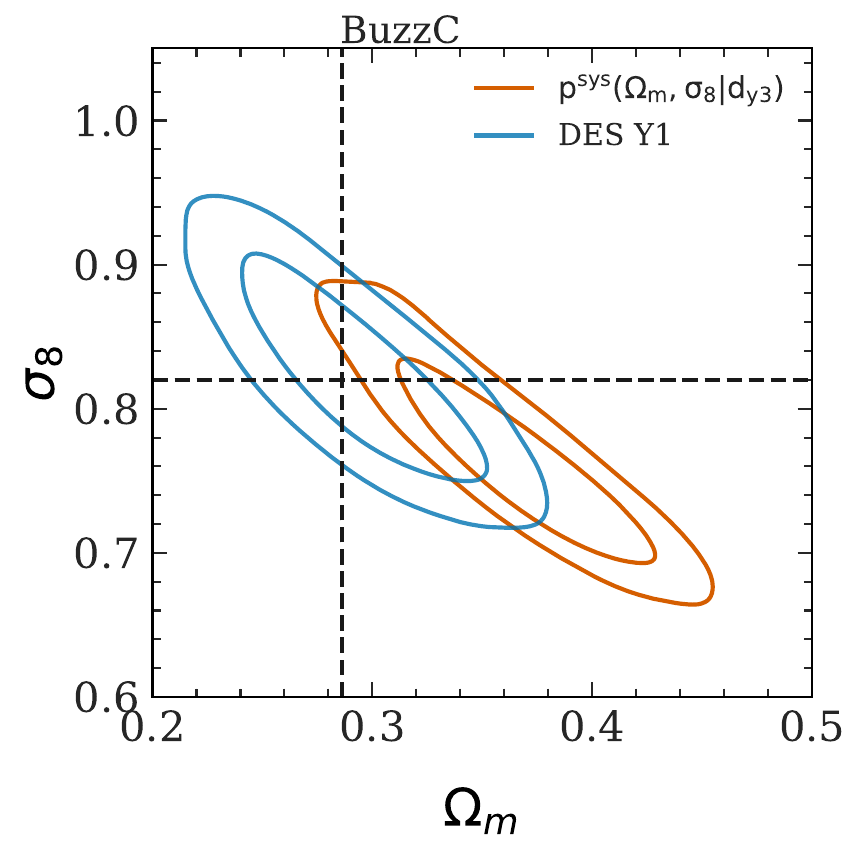}

\caption{Constraints on $\Omega_{\rm m}$ and $\sigma_8$ from cluster number counts and four two-point correlation functions (\redmagic{} auto-correlations, \redmapper{}-\redmagic{} cross-correlations, \redmapper{} auto-correlations, and cluster lensing) measured in three versions of the \textsc{Buzzard} mock catalogs: \buzzarda{} (top), \buzzardb{} (lower left), and \buzzardc{} (lower right). Contours show the $68$ per cent and $95$ per cent confidence levels. The differences between these simulations are summarized in Table~\ref{tab:sims}. In each panel, {\it gray} contours show the constraints from individual realizations and the {\it orange} contours show the combination of these posteriors (equation \ref{eq:systematic}). The {\it blue} contours show the expected DES Y1 constraints shifted to center on the input cosmology of the simulation. The black dashed lines indicate the true cosmology, i.e. the input cosmology used to generate the simulation. This indicates that our method is unbiased at the accuracy level needed for DES Y1 data.
}
\label{fig:countour_main}
\end{figure*}

\subsection{Fiducial cosmological parameter constraints}
\label{sec:simmain}

We test whether our pipeline can correctly recover the cosmological parameters in simulated data following the method developed in \cite{Niall}.
We assume that the 
potential systematically biased posterior on parameters $\bmath{\theta}$,  $P^{\rm{sys}}(\bmath{\theta}, s_i)$ inferred from analyzing a simulated data vector $s_i$ can be related to the true posterior $P(\bmath{\theta}|s_i)$ by a shift $\Delta \bmath{\theta}$ in the parameter space:
\begin{equation}
\label{eq:sysdef}
    P^{\rm{sys}}\left(\bmath{\theta}, s_i\right) = P\left(\bmath{\theta}-\Delta \bmath{\theta}|s_i\right).
\end{equation}
To quantify the significance and the size of potential systematics we estimate the posterior of $\Delta \bmath{\theta}$ by analyzing a set of simulated data vectors $\{s_i\}$, %
all generated from the same true cosmological parameters $\bmath{\theta}_{\mathrm{true}}$. In the following, we use Bayes' theorem to relate the posterior of $\Delta \bmath{\theta}$ ($P(\Delta \bmath{\theta}| \{s_i\}, \bmath{\theta}_{\mathrm{true}})$) to the potential systematically biased posterior ($P^{\rm{sys}}(\bmath{\theta}, s_i)$) inferred from analyzing individual simulated data vectors:
\begin{align}
\label{eq:systematic}
    P\left(\Delta \bmath{\theta}| \{s_i\}, \bmath{\theta}_{\rm{true}}\right) &=\frac{P\left ( \{s_i\}| \Delta \bmath{\theta}, \bmath{\theta}_{\rm{true}}\right)P\left(\Delta \bmath{\theta}\right)}{P\left(\{s_i\}\right)} \nonumber\\
    &= \frac{P\left(\Delta \bmath{\theta}\right)\prod_{i=1}^N P\left(s_i| \Delta \bmath{\theta}, \bmath{\theta}_{\rm{true}}\right)}{\prod_{i=1}^N P\left(s_i\right)} \nonumber\\
    &\propto \frac{\prod_{i=1}^N P\left(s_i|\bmath{\theta}_{\rm{true}}\right)}{\prod_{i=1}^N P\left(s_i\right)} \nonumber\\
     &\propto \prod_{i=1}^N P\left(\bmath{\theta}_{\rm{true}}|s_i\right)\nonumber\\
      &\propto \prod_{i=1}^N P^{\rm{sys}}\left(\bmath{\theta}_{\rm{true}}+\Delta \bmath{\theta}|s_i\right)\,, 
\end{align}
where we assume that we analyze N simulated universes. 
In the second equality, we assume that realizations of the simulated universe are mutually independent; in the third equality, we assume that the parameter shift ($\Delta \bmath{\theta}$) is drawn from a flat prior and the generation of simulated universe only depends on the true cosmological parameters $\bmath{\theta}_{\rm{true}}$; in the fifth equality, we substitute equation \ref{eq:sysdef}. From equation \ref{eq:systematic}, we can estimate the probability of systematic biases on a parameter by computing the product of the parameter posteriors from analyzing each simulation realization. 

In this paper, we focus on estimating systematic biases in $\Omega_{\rm m}$ and $\sigma_8$, the two cosmological parameters we expect to be well constrained by analyzing the DES-Y1 data.  
Fig.~\ref{fig:countour_main} shows $68$ per cent and $95$ per cent constraints on $\Omega_{\rm m}$ and $\sigma_8$ from the three versions of the \textsc{Buzzard} mock catalogs, marginalizing over 16 nuisance parameters and 3 cosmological parameters as described in section~\ref{sec:analysis choice}. The {\it gray} contours show the constraints from individual realizations and the {\it orange} contours represent the combination of constraints from all realizations. The black dashed lines indicate the true cosmology of the \textsc{Buzzard} mock catalogs. %
For comparison, in all panels in Fig.~\ref{fig:countour_main}, the {\it blue} contours denote the expected $68$ per cent and $95$ per cent constraints from the DES-Y1 data, which are estimated from analyzing noiseless theory data vectors with the covariance matrix generated based on DES-Y1 data \citep{DESY1KP, Lindsy}. %

From equation \ref{eq:systematic},  the {\it orange} contours can be related to the posterior of systematics ($P\left(\Delta \bmath{\theta}| \{s_i\}, \bmath{\theta}_{\rm{true}}\right)$) on $\Omega_{\rm m}$ and $\sigma_8$. Given the posterior of systematics, we can caluclate what is the minimal confidence interval containing the null hypothesis ($\Delta \theta=0$). We find that the $96.9$, $93.0$, $99.2$ per cent confidence intervals contain $\Delta \theta=0$ for \buzzarda{}, \buzzardb{}, and \buzzardc{} respectively. 
These results indicate the presence of biases in our simulation results at a $\approx 95$ per cent confidence.  We note, however, that the statistical power of the simulations is 10 times that of DES Y1 for \buzzarda\ and \buzzardb, and 3 times that of DES Y1 for \buzzardc.
To determine whether this level of systematic is important for a DES Y1 expeirment, we calculate the probability that the systematic shift in $\sigma_8$ and $\Omega_{\rm m}$ is contained within the $68$ per cent likelihood contour of the DES Y1 experiment. 
Roughly speaking, this is the probability that systematic shifts in the parameters are smaller than the statistical uncertainties. We find $P\left(\rm{sys} < \sigma_{\rm{Y1}}\right) = 0.65$ for \buzzarda{}, $P\left(\rm{sys} < \sigma_{\rm{Y1}}\right) = 0.57$ for \buzzardb{}, and $P\left(\rm{sys} < \sigma_{\rm{Y1}}\right) = 0.19$ for \buzzardc{}. Our analysis are %
consistent with negligible to modest parameter biases.

To decide whether enlarging the error of the analysis on real data is needed to accommodate this systematic, we need to understand whether this systematic is due to flaws in analysis pipeline or other sources, such as statistical fluctuations or flaws in the simulations. We therefore perform an analysis %
combining galaxy--galaxy lensing and galaxy clustering (2$\times$2pt analysis) on the same set of simulations. %
We find a similar level of bias %
for $\Omega_{\rm{m}}$ and $\sigma_8$ (see appendix~\ref{fig:app:2x2pt_clustercomparison} for details). From this analysis, we believe that the modest parameter biases found in Figure~\ref{fig:countour_main}  are not due to flaws in the analysis pipeline, but rather reflect an unlucky draw and/or possible flaws in the simulations that impact multiple large-scale structure analyses in similar ways.  For this reason, %
we decide not to enlarge the error bar on the analysis of real data.  More simulations are required to understand the sources of this parameter bias, and we plan to increase the number of simulations in future work. We also caution that we are not able to combine the three versions of the \textsc{Buzzard} simulations to arrive at a stronger statement because they share the same underlying matter distribution, and are therefore not mutually independent realizations.%

We summarize the estimation of systematic biases from simulations in Table~\ref{tab:mainsum}.

\begin{figure}
\includegraphics[width=0.5\textwidth]{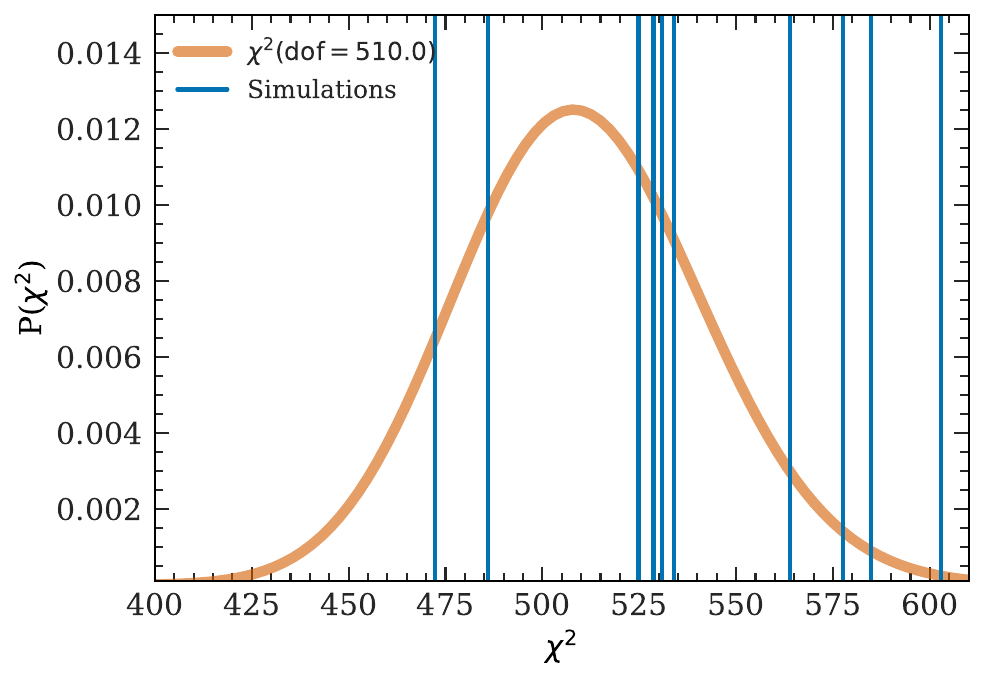}\hspace{-0.05\textwidth}
\caption{Comparison of the distribution of $\chi^2$s at the best-fit values in each realizations of \buzzarda{} (the {\it blue} vertical lines) and the expected distribution based on the number of degrees of freedom (the {\it orange} line). The agreement between the {\it orange} line and the distribution of {\it blue} lines indicates the validity of the theory covariance matrix.
}
\label{fig:chi2}
\end{figure}

\subsection{Performance of the theory covariance matrix}
\label{sec:cov}

One of the key components of this paper is the computation of the theory covariance matrix, which is used to combine the cluster abundance measurements and two-point statistics. In this section, we validate the performance of the covariance matrix using the \textsc{Buzzard} mock catalogs. Because we do not expect the performance of the theoretical covariance matrix %
to depend on the details of galaxy--halo connection, we focus on \buzzarda{} in this section. %
In each realizations of \buzzarda{}, we use Nelder-Mead algorithm \citep{journals/coap/GaoH12} to find the best-fit parameters. We then compute the distribution of $\chi^2$s at the best-fit parameters recovered from the 10 independent realizations, %
and compared it to %
the $\chi^2$ distribution with the appropriate %
number of degrees of freedom. The number of degrees of freedom is set to the number of entries in the data vector%
minus the number of parameters that %
are constrained by the data. Fig.~\ref{fig:chi2} shows this comparison.
To quantify the difference between these two distributions, we perform the Kolmogorov--Smirnov (K--S) test.
The resulting $p$-value is 1.1 per cent, indicating that the two probability distributions are consistent with one another.  The low probability may suggest the presence of unmodelled uncertainties, though confirming the existence of such errors will require increasing the number of simulations analyzed.

\section{Conclusions}
\label{sec:conclusion}

Combined-probe analyses have been demonstrated to be a powerful %
cosmological inferences tools %
\citep{DESY1KP, KIDScombine,2018MNRAS.474.4894J}. Not only do they allow the extraction of more cosmological information than that accessible to individual probes, but %
these analyses also provide an opportunity to internally calibrate possible systematics. Despite the promise, combined-probe analyses face many challenges, 
particularly with regards to the necessity of theoretical assumptions and the need for reliable covariance matrix estimates.
Thus, it is important to systematically validate methods %
of any new combined-probe analysis.  

In this work, we develop and validate %
a  method of combining the abundance of galaxy clusters %
with four two-point statistics: %
galaxy clustering ($w_{gg}$), galaxy--galaxy cluster cross-correlations ($w_{cg}$), galaxy cluster auto-correlations ($w_{cc}$), and cluster lensing ($\gamma_{t,c}$). %
Our methodology is validated using three versions of the \textsc{Buzzard} mock catalogs \citep[][summarized in Table~\ref{tab:sims}]{JoeBuzzard} with realistic galaxy and galaxy cluster selections. Based on this simulation analysis, we identify %
a boost in the clustering amplitude of \redmapper{} galaxy clusters due to %
selection effects in \redmapper.
Specifically, we find that \redmapper\ clusters 
in simulations are impacted by contamination with structures along the line of sight (projection effects), and that they are preferentially aligned along the line of sight.
\cite{Tomomi} find similar biases in cluster lensing and cluster auto-correlations based on their mock \redmapper{} catalogs. While the performance of mock \redmapper{} catalogs is arguably less sensitive
to %
the complicated galaxy--halo connection model, their mock \redmapper{} algorithm is not the same and will not include all of the same systematics as the full \redmapper{} algorithm being run on the data and on the simulations herein.
The three versions of the \textsc{Buzzard} mock catalogs that are used to validate our pipeline include many of the complexities %
in the real data: varying magnitude errors due to survey depth variations affect galaxy selection; correlations between mass tracers and the cluster selection efficiency %
impact the distribution of galaxy cluster samples; photometric redshift estimations of source galaxies are implemented; and realistic galaxy--halo connection models that are more sophisticated than the developed theoretical model are considered. 
Our analyses of these synthetic data sets are the first end-to-end tests of a cluster cosmology pipeline on realistic simulated galaxy data sets.

Our main results and conclusions can be summarized as follows:
\begin{enumerate}
    \item We identify an additional large-scale cluster bias due to \redmapper{} selection. We find that this selection bias can be well explained by projection effects and halo orientation biases, the two known systematics affecting the weak lensing signal of optically selected galaxy clusters (\citetalias{DES_cluster_cosmology}; \citealt{Heidiorientation}).
    \item We develop a model and a theory covariance matrix to combine the galaxy cluster abundance and two-point statistics (galaxy clustering, galaxy--cluster cross-correlations, cluster auto-correlations, and cluster lensing).
    \item We validate the model by analyzing three versions of the \textsc{Buzzard} mock catalogs. %
    Our simulation analysis is statistically consistent with no systematics biases, though there is $\approx 2\sigma$ evidence for a bias comparable to the statistical uncertainties in DES Y1.  We argue in Appendix \ref{app:2x2pt_4x2pt} that this bias is unlikely due to flaws in the analysis pipeline but rather due to unfortunate draws and/or flaws in the simulations that impact other large-scale structure probes similarly.  A definitive conclusion must await simulations with more constraining power. %
    \item We validate the theory covariance matrix by comparing the distribution of $\chi^2$s at the best-fit parameters from analyses of each realization of the \textsc{Buzzard} mock catalogs and the expected $\chi^2$ distributions. We perform the Kolmogorov-Smirnov (K-S) test and find the $p$-value is $1.1$ per cent, indicating that the two probability distributions are consistent with one another. %
    \item We stress test the analysis pipeline by analyzing theory data vectors contaminated by systematics that are not fully captured by the \textsc{Buzzard} mock catalogs: lowering the cluster lensing one-halo term in the lowest richness bin by $50$ per cent, adding a non-linear clustering term, and using a more complicated functional form of the richness--mass relation of \redmapper{} clusters. We find that the inference pipeline is robust against these possible systematics. 
\end{enumerate}
In the near future, we plan to apply the pipeline developed in this paper %
to the DES-Y1 data set. The covariance matrix developed here enables us to combine our results with the constraints from the 3$\times$2pt DES Y1 key project \citepalias{DESY1KP}. Although we have been focusing on the application on cosmological constraints from optically identified galaxy clusters, we note that there are many potential applications.  First, the comparison of different two-point correlations can shed additional light on systematics of galaxy clusters as a cosmological probe. For example, for cluster cosmology analyses using large-scale information, we can test the robustness of bias models by comparing different two-point correlations. Second, the same pipeline can be applied to galaxy clusters selected in other wavelengths, such as X-ray and microwave. We forecast these possibilities in a companion paper \citep*{forecastpaper}. %
Our results suggest that despite the surprising results of the DES Y1 cluster abundance analysis \citepalias{DES_cluster_cosmology}, a multi-probe cluster cosmology approach based on photometrically selected samples may recover unbiased cosmological parameter information when restricting the analysis to large scales only.  Moreover, because our analysis is especially well suited for being combined with the now popular 3$\times$2pt analysis (combining cosmic shear, galaxy--galaxy lensing, and galaxy clustering), we expect the approach highlighted here may become the standard for the cosmological analysis of near-future photometric cluster samples.

\section*{Acknowledgements}
\label{sec:acknowledgements}
We thank Sebastian Bocquet, August Evrard, Oliver Friedrich, and Xiao Fang for helpful discussions and comments on the manuscript. We thank Vivian Miranda for early implementations of parts of the code in this study. The original idea was discussed during the Lighthouse workshop, organized by DG, EK, and Adam Mantz, held March 2017 at Point Montara. 
We thank the workshop participants for early discussions, and the KIPAC workshop program for support.  This paper has gone through internal review by the DES collaboration. This work was supported in part by the U.S. Department of Energy contract to SLAC no. DE-AC02- 76SF00515 (CH, DG, RW).  CH and EK are supported in part by NASA ROSES ATP 16-ATP16-0084. EK is supported in part by Department of Energy grant DE-SC0020247.  ER is supported by DOE grants DE-SC0015975 and DE-SC0009913, and by NSF Grant AST-2009401.  ER also acknowledges funding from the Cottrell Scholar program of the Research Corporation for Science Advancement. HW is supported by NSF Grant AST-1516997. This work was supported by the Department of Energy, Laboratory Directed Research and Development program at SLAC National Accelerator Laboratory, under contract DE-AC02-76SF00515 and as part of the Panofsky Fellowship awarded to DG. Some of the computing for this project was performed on the Sherlock cluster; we thank Stanford University and the Stanford Research Computing Center for providing computational resources and support that contributed to these results. 

Funding for the DES Projects has been provided by the U.S. Department of Energy, the U.S. National Science Foundation, the Ministry of Science and Education of Spain, 
the Science and Technology Facilities Council of the United Kingdom, the Higher Education Funding Council for England, the National Center for Supercomputing 
Applications at the University of Illinois at Urbana-Champaign, the Kavli Institute of Cosmological Physics at the University of Chicago, 
the Center for Cosmology and Astro-Particle Physics at the Ohio State University,
the Mitchell Institute for Fundamental Physics and Astronomy at Texas A\&M University, Financiadora de Estudos e Projetos, 
Funda{\c c}{\~a}o Carlos Chagas Filho de Amparo {\`a} Pesquisa do Estado do Rio de Janeiro, Conselho Nacional de Desenvolvimento Cient{\'i}fico e Tecnol{\'o}gico and 
the Minist{\'e}rio da Ci{\^e}ncia, Tecnologia e Inova{\c c}{\~a}o, the Deutsche Forschungsgemeinschaft and the Collaborating Institutions in the Dark Energy Survey. 

The Collaborating Institutions are Argonne National Laboratory, the University of California at Santa Cruz, the University of Cambridge, Centro de Investigaciones Energ{\'e}ticas, 
Medioambientales y Tecnol{\'o}gicas-Madrid, the University of Chicago, University College London, the DES-Brazil Consortium, the University of Edinburgh, 
the Eidgen{\"o}ssische Technische Hochschule (ETH) Z{\"u}rich, 
Fermi National Accelerator Laboratory, the University of Illinois at Urbana-Champaign, the Institut de Ci{\`e}ncies de l'Espai (IEEC/CSIC), 
the Institut de F{\'i}sica d'Altes Energies, Lawrence Berkeley National Laboratory, the Ludwig-Maximilians Universit{\"a}t M{\"u}nchen and the associated Excellence Cluster Universe, 
the University of Michigan, NFS's NOIRLab, the University of Nottingham, The Ohio State University, the University of Pennsylvania, the University of Portsmouth, 
SLAC National Accelerator Laboratory, Stanford University, the University of Sussex, Texas A\&M University, and the OzDES Membership Consortium.

Based in part on observations at Cerro Tololo Inter-American Observatory at NSF's NOIRLab (NOIRLab Prop. ID 2012B-0001; PI: J. Frieman), which is managed by the Association of Universities for Research in Astronomy (AURA) under a cooperative agreement with the National Science Foundation.

The DES data management system is supported by the National Science Foundation under Grant Numbers AST-1138766 and AST-1536171.
The DES participants from Spanish institutions are partially supported by MICINN under grants ESP2017-89838, PGC2018-094773, PGC2018-102021, SEV-2016-0588, SEV-2016-0597, and MDM-2015-0509, some of which include ERDF funds from the European Union. IFAE is partially funded by the CERCA program of the Generalitat de Catalunya.
Research leading to these results has received funding from the European Research
Council under the European Union's Seventh Framework Program (FP7/2007-2013) including ERC grant agreements 240672, 291329, and 306478.
We  acknowledge support from the Brazilian Instituto Nacional de Ci\^encia
e Tecnologia (INCT) do e-Universo (CNPq grant 465376/2014-2).

This manuscript has been authored by Fermi Research Alliance, LLC under Contract No. DE-AC02-07CH11359 with the U.S. Department of Energy, Office of Science, Office of High Energy Physics.

\section*{DATA AVAILABILITY}
The simulation data underlying this article will be shared on reasonable request to the corresponding author.

\bibliographystyle{mn2e_2author_arxiv_amp.bst}
\bibliography{sample.bib} %

\begin{thebibliography}{71}
\providecommand{\natexlab}[1]{#1}
\providecommand{\url}[1]{\texttt{#1}}
\providecommand{\urlprefix}{URL }
\providecommand{\eprint}[1][]{\url{#1}}

\bibitem[{{Abbott} et~al.(2018){Abbott} \& {Abdalla} et~al.}]{DESY1KP}
{Abbott}, T.~M.~C., {Abdalla}, F.~B., {Alarcon}, A., et~al., 2018, \prd, 98, 4,
  043526

\bibitem[{{Abbott} et~al.(2020){Abbott} \& {Aguena}
  et~al.}]{DES_cluster_cosmology}
{Abbott}, T.~M.~C., {Aguena}, M., {Alarcon}, A., et~al., 2020, \prd, 102, 2,
  023509

\bibitem[{{Allen} et~al.(2011){Allen} \& {Evrard} \&
  {Mantz}}]{2011ARA&A..49..409A}
{Allen}, S.~W., {Evrard}, A.~E., {Mantz}, A.~B., 2011, \araa, 49, 1, 409

\bibitem[{{Baldauf} et~al.(2012){Baldauf} \& {Seljak} \& {Desjacques} \&
  {McDonald}}]{2012PhRvD..86h3540B}
{Baldauf}, T., {Seljak}, U., {Desjacques}, V., {McDonald}, P., 2012, \prd, 86,
  8, 083540

\bibitem[{{Becker}(2013)}]{2013MNRAS.435..115B}
{Becker}, M.~R., 2013, \mnras, 435, 1, 115

\bibitem[{{Ben{\'\i}tez}(2000)}]{2000ApJ...536..571B}
{Ben{\'\i}tez}, N., 2000, \apj, 536, 2, 571

\bibitem[{{Bhattacharya} et~al.(2013){Bhattacharya} \& {Habib} \& {Heitmann} \&
  {Vikhlinin}}]{2013ApJ...766...32B}
{Bhattacharya}, S., {Habib}, S., {Heitmann}, K., {Vikhlinin}, A., 2013, \apj,
  766, 1, 32

\bibitem[{{Blas} et~al.(2011){Blas} \& {Lesgourgues} \& {Tram}}]{CLASS}
{Blas}, D., {Lesgourgues}, J., {Tram}, T., 2011, \jcap, 2011, 7, 034

\bibitem[{{Bleem} et~al.(2020){Bleem} \& {Bocquet} et~al.}]{Lindsy}
{Bleem}, L.~E., {Bocquet}, S., {Stalder}, B., et~al., 2020, \apjs, 247, 1, 25

\bibitem[{{Carron}(2013)}]{2013A&A...551A..88C}
{Carron}, J., 2013, \aap, 551, A88

\bibitem[{{Cawthon} et~al.(2018){Cawthon} \& {Davis}
  et~al.}]{2018MNRAS.481.2427C}
{Cawthon}, R., {Davis}, C., {Gatti}, M., et~al., 2018, \mnras, 481, 2, 2427

\bibitem[{{Chisari} et~al.(2019){Chisari} \& {Alonso} et~al.}]{CCL}
{Chisari}, N.~E., {Alonso}, D., {Krause}, E., et~al., 2019, \apjs, 242, 1, 2

\bibitem[{{Chisari} \& {Pontzen}(2019)}]{2019PhRvD.100b3543C}
{Chisari}, N.~E., {Pontzen}, A., 2019, \prd, 100, 2, 023543

\bibitem[{{Cooray} \& {Sheth}(2002)}]{2002PhR...372....1C}
{Cooray}, A., {Sheth}, R., 2002, \physrep, 372, 1, 1

\bibitem[{{Costanzi} et~al.(2019{\natexlab{a}}){Costanzi} \& {Rozo}
  et~al.}]{Matteoprojection}
{Costanzi}, M., {Rozo}, E., {Rykoff}, E.~S., et~al., 2019{\natexlab{a}},
  \mnras, 482, 1, 490

\bibitem[{{Costanzi} et~al.(2019{\natexlab{b}}){Costanzi} \& {Rozo}
  et~al.}]{matteoSDSS}
{Costanzi}, M., {Rozo}, E., {Simet}, M., et~al., 2019{\natexlab{b}}, \mnras,
  488, 4, 4779

\bibitem[{{DeRose} et~al.(2019){DeRose} \& {Wechsler} et~al.}]{JoeBuzzard}
{DeRose}, J., {Wechsler}, R.~H., {Becker}, M.~R., et~al., 2019, arXiv e-prints,
  arXiv:1901.02401

\bibitem[{{DeRose} et~al.(2020)}]{campaper}
{DeRose}, J., et~al., 2020, in prep.

\bibitem[{{DES Collaboration} et~al.(2020)}]{Y3Gold}
{DES Collaboration}, 2020, in prep.

\bibitem[{{Dietrich} et~al.(2014){Dietrich} \& {Zhang}
  et~al.}]{2014MNRAS.443.1713D}
{Dietrich}, J.~P., {Zhang}, Y., {Song}, J., et~al., 2014, \mnras, 443, 2, 1713

\bibitem[{{Dodelson} \& {Schneider}(2013)}]{2013PhRvD..88f3537D}
{Dodelson}, S., {Schneider}, M.~D., 2013, \prd, 88, 6, 063537

\bibitem[{{Eisenstein} \& {Hu}(1998)}]{1998ApJ...496..605E}
{Eisenstein}, D.~J., {Hu}, W., 1998, \apj, 496, 2, 605

\bibitem[{{Fang} et~al.(2020){Fang} \& {Krause} \& {Eifler} \&
  {MacCrann}}]{2019arXiv191111947F}
{Fang}, X., {Krause}, E., {Eifler}, T., {MacCrann}, N., 2020, \jcap, 2020, 5,
  010

\bibitem[{Gao \& Han(2012)}]{journals/coap/GaoH12}
Gao, F., Han, L., 2012, Comp. Opt. and Appl., 51, 1, 259

\bibitem[{{Hartlap} et~al.(2007){Hartlap} \& {Simon} \&
  {Schneider}}]{2007A&A...464..399H}
{Hartlap}, J., {Simon}, P., {Schneider}, P., 2007, \aap, 464, 1, 399

\bibitem[{{Hoyle} et~al.(2018){Hoyle} \& {Gruen} et~al.}]{2018MNRAS.478..592H}
{Hoyle}, B., {Gruen}, D., {Bernstein}, G.~M., et~al., 2018, \mnras, 478, 1, 592

\bibitem[{{Jarvis} et~al.(2004){Jarvis} \& {Bernstein} \&
  {Jain}}]{2004MNRAS.352..338J}
{Jarvis}, M., {Bernstein}, G., {Jain}, B., 2004, \mnras, 352, 1, 338

\bibitem[{{Joudaki} et~al.(2018{\natexlab{a}}){Joudaki} \& {Blake}
  et~al.}]{KIDS3x2pt1}
{Joudaki}, S., {Blake}, C., {Johnson}, A., et~al., 2018{\natexlab{a}}, \mnras,
  474, 4, 4894

\bibitem[{{Joudaki} et~al.(2018{\natexlab{b}}){Joudaki} \& {Blake}
  et~al.}]{2018MNRAS.474.4894J}
{Joudaki}, S., {Blake}, C., {Johnson}, A., et~al., 2018{\natexlab{b}}, \mnras,
  474, 4, 4894

\bibitem[{{Krause} \& {Eifler}(2017)}]{cosmolike2016}
{Krause}, E., {Eifler}, T., 2017, \mnras, 470, 2, 2100

\bibitem[{{Krause} et~al.(2017){Krause} \& {Eifler} et~al.}]{DESmethod}
{Krause}, E., {Eifler}, T.~F., {Zuntz}, J., et~al., 2017, arXiv e-prints,
  arXiv:1706.09359

\bibitem[{{Krause} et~al.(2020){Krause} \& {To} et~al.}]{forecastpaper}
{Krause}, E., {To}, C.-H., et~al., 2020, in prep.

\bibitem[{{Lacasa} \& {Rosenfeld}(2016)}]{2016JCAP...08..005L}
{Lacasa}, F., {Rosenfeld}, R., 2016, \jcap, 2016, 8, 005

\bibitem[{{Landy} \& {Szalay}(1993)}]{1993ApJ...412...64L}
{Landy}, S.~D., {Szalay}, A.~S., 1993, \apj, 412, 64

\bibitem[{{Lazeyras} et~al.(2016){Lazeyras} \& {Wagner} \& {Baldauf} \&
  {Schmidt}}]{2016JCAP...02..018L}
{Lazeyras}, T., {Wagner}, C., {Baldauf}, T., {Schmidt}, F., 2016, \jcap, 2016,
  2, 018

\bibitem[{{Leauthaud} et~al.(2017){Leauthaud} \& {Saito} et~al.}]{Lensingislow}
{Leauthaud}, A., {Saito}, S., {Hilbert}, S., et~al., 2017, \mnras, 467, 3, 3024

\bibitem[{{Lehmann} et~al.(2017){Lehmann} \& {Mao} \& {Becker} \& {Skillman} \&
  {Wechsler}}]{2017ApJ...834...37L}
{Lehmann}, B.~V., {Mao}, Y.-Y., {Becker}, M.~R., {Skillman}, S.~W., {Wechsler},
  R.~H., 2017, \apj, 834, 1, 37

\bibitem[{{Limber}(1953)}]{Limber}
{Limber}, D.~N., 1953, \apj, 117, 134

\bibitem[{{MacCrann} et~al.(2018){MacCrann} \& {DeRose} et~al.}]{Niall}
{MacCrann}, N., {DeRose}, J., {Wechsler}, R.~H., et~al., 2018, \mnras, 480, 4,
  4614

\bibitem[{{McClintock} et~al.(2019{\natexlab{a}}){McClintock} \& {Rozo}
  et~al.}]{biasemu}
{McClintock}, T., {Rozo}, E., {Banerjee}, A., et~al., 2019{\natexlab{a}}, arXiv
  e-prints, arXiv:1907.13167

\bibitem[{{McClintock} et~al.(2019{\natexlab{b}}){McClintock} \& {Rozo}
  et~al.}]{HMFemulator}
{McClintock}, T., {Rozo}, E., {Becker}, M.~R., et~al., 2019{\natexlab{b}},
  \apj, 872, 1, 53

\bibitem[{{McClintock} et~al.(2019{\natexlab{c}}){McClintock} \& {Varga}
  et~al.}]{Tomclusterlensing}
{McClintock}, T., {Varga}, T.~N., {Gruen}, D., et~al., 2019{\natexlab{c}},
  \mnras, 482, 1, 1352

\bibitem[{{McDonald} \& {Roy}(2009)}]{2009JCAP...08..020M}
{McDonald}, P., {Roy}, A., 2009, \jcap, 2009, 8, 020

\bibitem[{{McEwen} et~al.(2016){McEwen} \& {Fang} \& {Hirata} \&
  {Blazek}}]{2016JCAP...09..015M}
{McEwen}, J.~E., {Fang}, X., {Hirata}, C.~M., {Blazek}, J.~A., 2016, \jcap,
  2016, 9, 015

\bibitem[{{Nicola} et~al.(2020){Nicola} \& {Dunkley} \& {Spergel}}]{Andrina}
{Nicola}, A., {Dunkley}, J., {Spergel}, D.~N., 2020, arXiv e-prints,
  arXiv:2006.00008

\bibitem[{{Oguri} \& {Takada}(2011)}]{Ogori}
{Oguri}, M., {Takada}, M., 2011, \prd, 83, 2, 023008

\bibitem[{{Osato} et~al.(2018){Osato} \& {Nishimichi} \& {Oguri} \& {Takada} \&
  {Okumura}}]{2018MNRAS.477.2141O}
{Osato}, K., {Nishimichi}, T., {Oguri}, M., {Takada}, M., {Okumura}, T., 2018,
  \mnras, 477, 2, 2141

\bibitem[{{Rozo} et~al.(2016){Rozo} \& {Rykoff} et~al.}]{Redmagic}
{Rozo}, E., {Rykoff}, E.~S., {Abate}, A., et~al., 2016, \mnras, 461, 2, 1431

\bibitem[{{Rykoff} et~al.(2014){Rykoff} \& {Rozo} et~al.}]{Redmapper1}
{Rykoff}, E.~S., {Rozo}, E., {Busha}, M.~T., et~al., 2014, \apj, 785, 2, 104

\bibitem[{{Saito} et~al.(2014){Saito} \& {Baldauf} \& {Vlah} \& {Seljak} \&
  {Okumura} \& {McDonald}}]{2014PhRvD..90l3522S}
{Saito}, S., {Baldauf}, T., {Vlah}, Z., {Seljak}, U., {Okumura}, T.,
  {McDonald}, P., 2014, \prd, 90, 12, 123522

\bibitem[{{Salcedo} et~al.(2020){Salcedo} \& {Wibking}
  et~al.}]{2020MNRAS.491.3061S}
{Salcedo}, A.~N., {Wibking}, B.~D., {Weinberg}, D.~H., et~al., 2020, \mnras,
  491, 3, 3061

\bibitem[{{Schaan} et~al.(2014){Schaan} \& {Takada} \&
  {Spergel}}]{2014PhRvD..90l3523S}
{Schaan}, E., {Takada}, M., {Spergel}, D.~N., 2014, \prd, 90, 12, 123523

\bibitem[{{Sellentin} \& {Heavens}(2016)}]{2016MNRAS.456L.132S}
{Sellentin}, E., {Heavens}, A.~F., 2016, \mnras, 456, 1, L132

\bibitem[{{Sheldon} et~al.(2004){Sheldon} \& {Johnston}
  et~al.}]{2004AJ....127.2544S}
{Sheldon}, E.~S., {Johnston}, D.~E., {Frieman}, J.~A., et~al., 2004, \aj, 127,
  5, 2544

\bibitem[{{Sif{\'o}n} et~al.(2015){Sif{\'o}n} \& {Hoekstra}
  et~al.}]{2015A&A...575A..48S}
{Sif{\'o}n}, C., {Hoekstra}, H., {Cacciato}, M., et~al., 2015, \aap, 575, A48

\bibitem[{{Sinha} \& {Garrison}(2020)}]{2020MNRAS.491.3022S}
{Sinha}, M., {Garrison}, L.~H., 2020, \mnras, 491, 2, 3022

\bibitem[{{Sunayama} et~al.(2020){Sunayama} \& {Park} et~al.}]{Tomomi}
{Sunayama}, T., {Park}, Y., {Takada}, M., et~al., 2020, arXiv e-prints,
  arXiv:2002.03867

\bibitem[{{Takada} \& {Bridle}(2007)}]{Takada&Bridle}
{Takada}, M., {Bridle}, S., 2007, New Journal of Physics, 9, 12, 446

\bibitem[{{Takahashi} et~al.(2012){Takahashi} \& {Sato} \& {Nishimichi} \&
  {Taruya} \& {Oguri}}]{Takahashi2012}
{Takahashi}, R., {Sato}, M., {Nishimichi}, T., {Taruya}, A., {Oguri}, M., 2012,
  \apj, 761, 2, 152

\bibitem[{{Tinker} et~al.(2010){Tinker} \& {Robertson} et~al.}]{Tinker10}
{Tinker}, J.~L., {Robertson}, B.~E., {Kravtsov}, A.~V., et~al., 2010, \apj,
  724, 2, 878

\bibitem[{{To} et~al.(2019){To} \& {Reddick} \& {Rozo} \& {Rykoff} \&
  {Wechsler}}]{2019arXiv191001656T}
{To}, C.-H., {Reddick}, R.~M., {Rozo}, E., {Rykoff}, E., {Wechsler}, R.~H.,
  2019, arXiv e-prints, arXiv:1910.01656

\bibitem[{{van Uitert} et~al.(2018{\natexlab{a}}){van Uitert} \& {Joachimi}
  et~al.}]{KIDS3x2pt2}
{van Uitert}, E., {Joachimi}, B., {Joudaki}, S., et~al., 2018{\natexlab{a}},
  \mnras, 476, 4, 4662

\bibitem[{{van Uitert} et~al.(2018{\natexlab{b}}){van Uitert} \& {Joachimi}
  et~al.}]{KIDScombine}
{van Uitert}, E., {Joachimi}, B., {Joudaki}, S., et~al., 2018{\natexlab{b}},
  \mnras, 476, 4, 4662

\bibitem[{{Wechsler} et~al.(2020)}]{Addgal}
{Wechsler}, R., et~al., 2020, in prep.

\bibitem[{{Wechsler} \& {Tinker}(2018)}]{Risa2018}
{Wechsler}, R.~H., {Tinker}, J.~L., 2018, \araa, 56, 435

\bibitem[{{Weinberg} et~al.(2013){Weinberg} \& {Mortonson} \& {Eisenstein} \&
  {Hirata} \& {Riess} \& {Rozo}}]{2013PhR...530...87W}
{Weinberg}, D.~H., {Mortonson}, M.~J., {Eisenstein}, D.~J., {Hirata}, C.,
  {Riess}, A.~G., {Rozo}, E., 2013, \physrep, 530, 2, 87

\bibitem[{{Wu} et~al.(2019){Wu} \& {Weinberg} \& {Salcedo} \& {Wibking} \&
  {Zu}}]{HeidiCov}
{Wu}, H.-Y., {Weinberg}, D.~H., {Salcedo}, A.~N., {Wibking}, B.~D., {Zu}, Y.,
  2019, \mnras, 490, 2, 2606

\bibitem[{{Wu} et~al.(2020)}]{Heidiorientation}
{Wu}, H.-Y., et~al., 2020, in prep.

\bibitem[{{Zehavi} et~al.(2011){Zehavi} \& {Zheng}
  et~al.}]{2011ApJ...736...59Z}
{Zehavi}, I., {Zheng}, Z., {Weinberg}, D.~H., et~al., 2011, \apj, 736, 1, 59

\bibitem[{{Zhang} et~al.(2019){Zhang} \& {Jeltema} et~al.}]{miscentering}
{Zhang}, Y., {Jeltema}, T., {Hollowood}, D.~L., et~al., 2019, \mnras, 487, 2,
  2578

\bibitem[{{Zuntz} et~al.(2018){Zuntz} \& {Sheldon}
  et~al.}]{2018MNRAS.481.1149Z}
{Zuntz}, J., {Sheldon}, E., {Samuroff}, S., et~al., 2018, \mnras, 481, 1, 1149

\end{thebibliography}

\appendix

\section{Investigation of \textsc{Buzzard} mock catalogs}
\label{app:simulationrealization}
\begin{figure}
\label{fig:app:debug}
\includegraphics[width=0.4\textwidth]{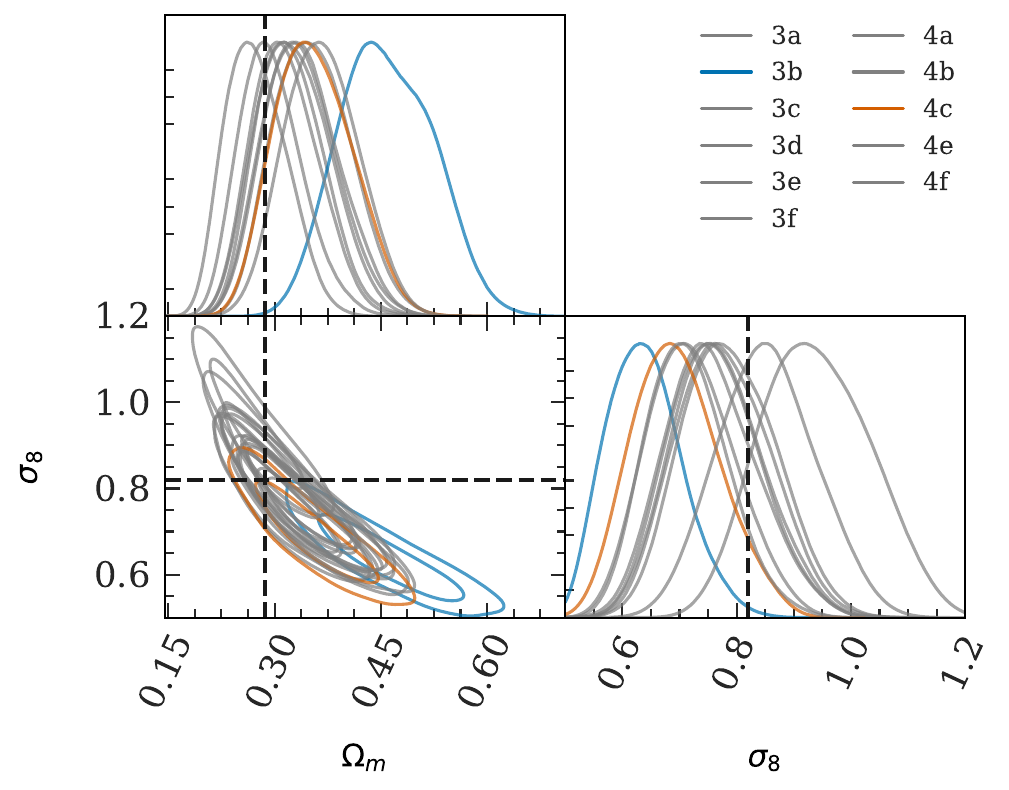}
\label{fig:app_2x2pt}
\includegraphics[width=0.4\textwidth]{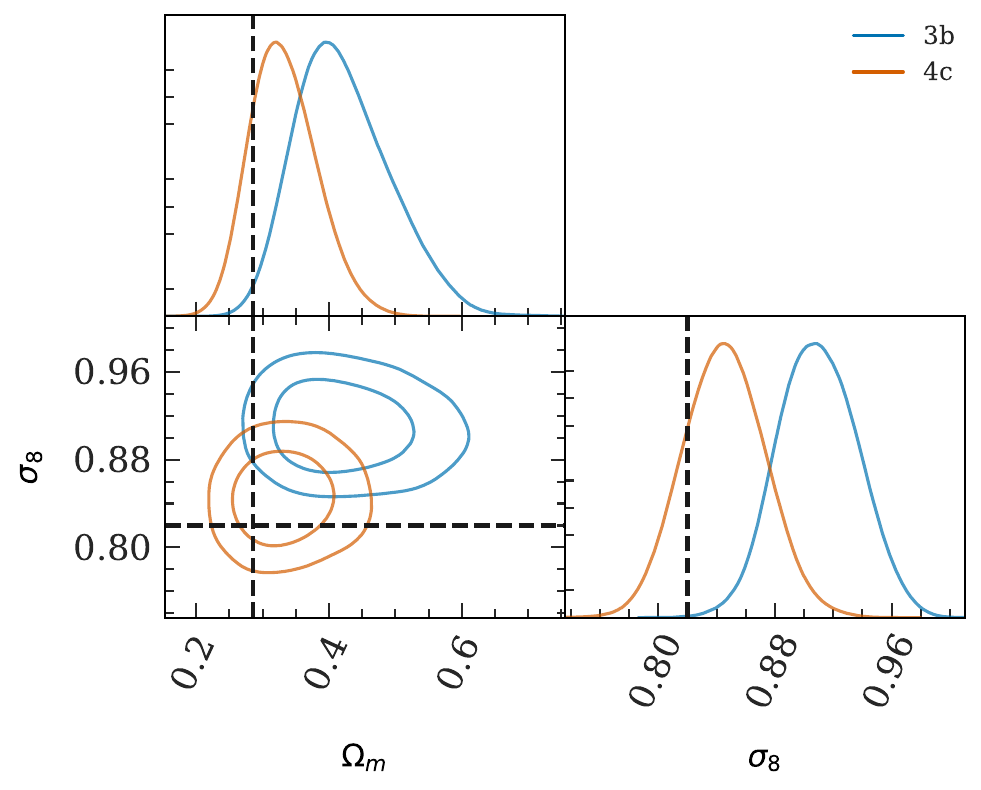}
\label{fig:app_wgg}
\includegraphics[width=0.4\textwidth]{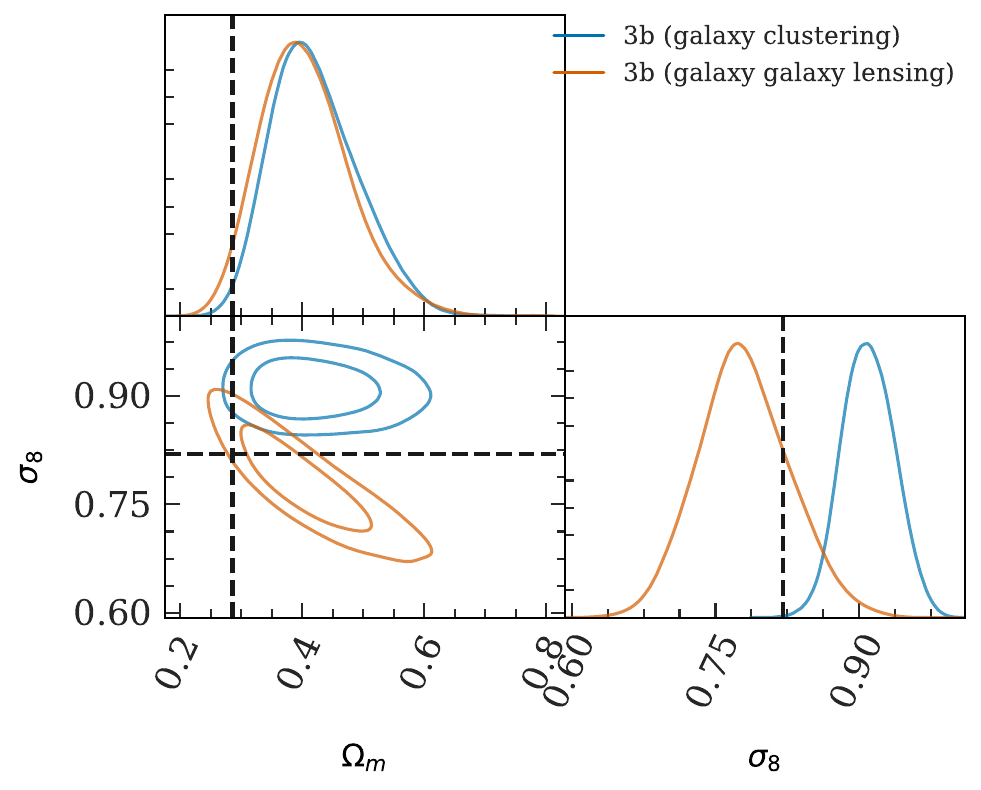}
\label{fig:app_wggl}
\caption{\textit{Top panel}: Constraints on $\Omega_{\rm m}$ and $\sigma_8$ from galaxy clustering and galaxy--galaxy lensing (2$\times$2pt) measured in each realizations in \buzzarda{}; contours show the $68$ per cent and $95$ per cent confidence levels. The \textit{ blue} contours denote the most biased realization and the \textit{ orange} contours denote the second most biased realization.
\textit{ Middle panel}: Constraints on $\Omega_{\rm m}$ and $\sigma_8$ from galaxy clustering; contours show the $68$ per cent and $95$ per cent confidence levels. In this analysis, we fix the galaxy biases at galaxy biases measured from cross-correlations of \redmagic{} galaxies and dark matter particles in each realization. Clearly, the figure shows that the galaxy clustering in realization 3b is biased. The same bias is not shown in the second most biased realization. This finding makes us believe that the galaxy clustering behaves differently in realization 3b from other realizations. 
\textit{ Bottom panel}: Constraints on $\Omega_{\rm m}$ and $\sigma_8$ from galaxy clustering and galaxy--galaxy lensing in realization 3b; contours show the $68$ per cent and $95$ per cent confidence levels. In this analysis, we fix the galaxy biases at galaxy biases measured from cross-correlations of \redmagic{} galaxies and dark matter particles. Clearly, the figure shows that the galaxy clustering is in tension with the galaxy--galaxy lensing in realization 3b.
}
\end{figure}

During the analysis, we find one realization being different than other realizations. In the specific realization (realization 3b of \buzzard{} and \buzzardb{}), we find that the combination of galaxy clustering and galaxy--galaxy lensing recover cosmological parameters that are  $3 \sigma$ away from the true cosmology. The top panel of Fig.~\ref{fig:app_2x2pt} shows the $68$ per cent and $95$ per cent constraints from the combination of galaxy clustering and galaxy--galaxy lensing. We can see that realization 3b is clearly an outlier compares to other realizations. Since the combination of galaxy clustering and galaxy--galaxy lensing pipeline has been thoroughly validated \citep*{Niall, DESmethod}, we interpret this as an indication that galaxies behave differently in realization 3b. We are then interested in  whether it is the galaxy clustering or galaxy--galaxy lensing data vector that causes this bias, because only galaxy clustering goes into the data vector in the analysis of this paper. We analyze the galaxy clustering data vector alone by fixing the galaxy biases at galaxy biases measured from cross-correlations of \redmagic{} galaxies and dark matter particles. We find that the galaxy clustering in realization 3b is problematic. The recovered cosmology is more than $3\sigma$ away from the truth. As a comparison, we show the galaxy clustering constraints in realization 4c, the second-most biased realization. The bias of cosmological parameters in realization 3b does not present in realization 4c. This finding indicates that the galaxy clustering behaves differently in realization 3b than other realizations. Further, as shown in the bottom panel of Fig.~\ref{fig:app_2x2pt}, we find a 3 $\sigma$ tension between galaxy clustering and galaxy--galaxy lensing in realization 3b, consolidating our conclusions that galaxy clustering in realization 3b is problematic. Although it is not included in our main analysis, we show the result of including this problematic realization in table~\ref{tab:mainsum_app}. 

\begin{table*}
\begin{tabular}{llll}
\hline
Simulations   & \buzzarda{} &\buzzardb{} \\ \hline
$\Delta \Omega_{\rm m} $ & $0.051 \pm 0.016$
&$0.051 \pm 0.016$\\
$\Delta \sigma_8 $ & $-0.077 \pm 0.020$ &
$-0.067 \pm 0.023$ \\
\hline
 minimal confidence interval encompasses $\Delta \theta=0$& 1.000
& 0.997\\
\hline
$P(\rm{sys}<\sigma_{\rm{Y1}})$ & 0.28 & 0.50  \\ 
$P(\rm{sys}<2\sigma_{\rm{Y1}})$  & 0.88 & 0.94  \\
 \hline
\end{tabular}
\caption{This table is the same as Table \ref{tab:mainsum} but includes realization 3b, a realization known to be problematic as shown in appendix~\ref{app:simulationrealization}.}
\label{tab:mainsum_app}
\end{table*}

\section{Comparison of the Halorun and the Full run}
\label{app:halorun_fullrun}
\begin{figure}
\centering
\includegraphics[width=0.5\textwidth]{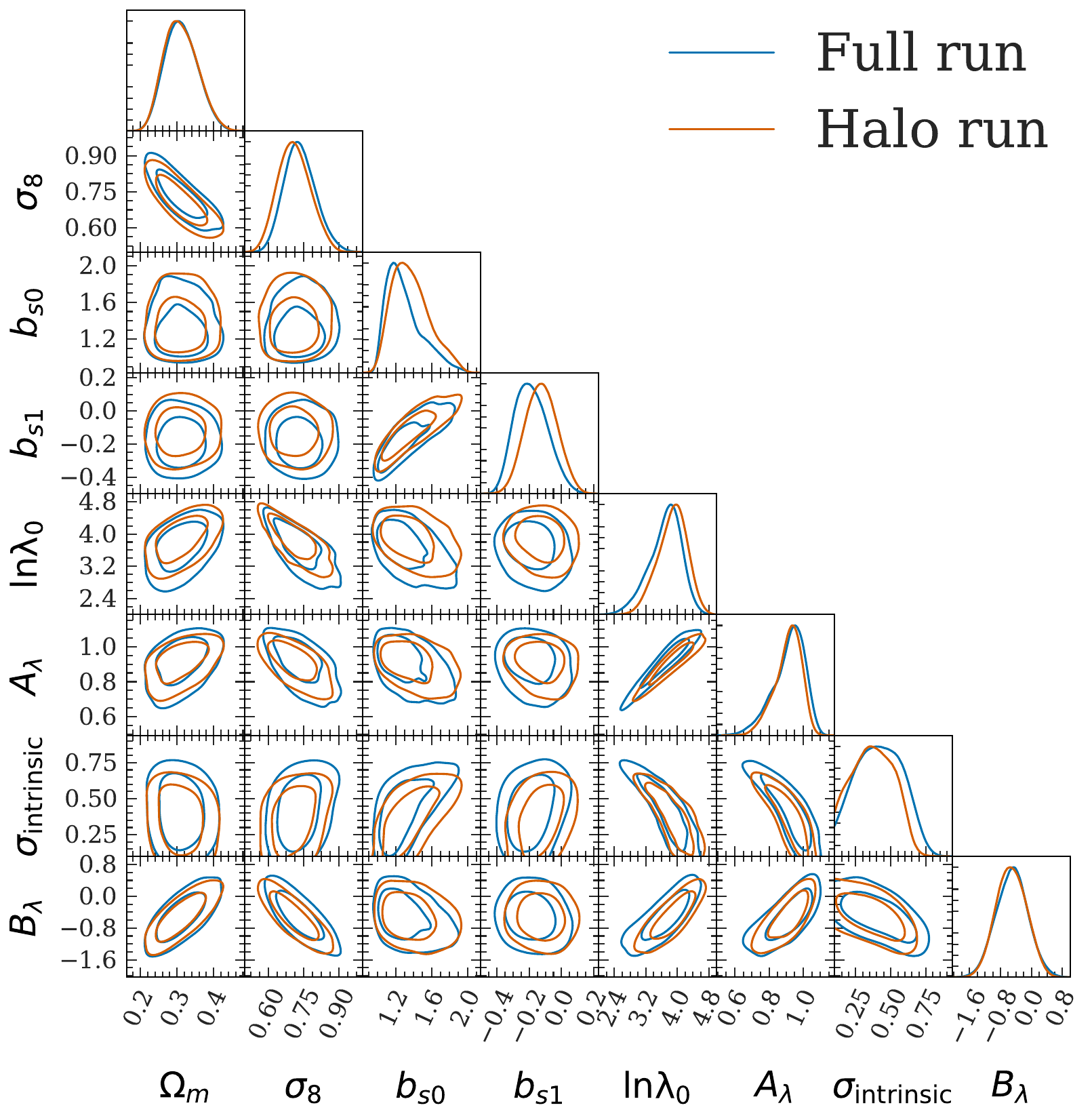}\hspace{-0.05\textwidth}
\caption{Constraints on cosmological parameters ($\Omega_{\rm m}$ and $\sigma_8$), richness--mass relation parameters ($\rm{ln}\lambda_0$, $A_\lambda$, $\sigma_{\rm{intrinsic}}$, $B_\lambda$), and selection bias parameters ($b_{\rm{s0}}$, $b_{\rm{s1}}$); contours show the $68$ per cent and $95$ per cent confidence levels.  Both colors are analysis on data vectors generated from the same \textsc{Buzzard} realization.  The {\it orange} lines are analysis on the \redmapper{} Halorun, in which \redmapper{} is run fixing the cluster centers at the halo centers. The {\it blue} lines indicate an analysis on the \redmapper{} Fullrun, where \redmapper{} is run with the same setting as the run on real data. The agreement between the {\it orange} and {\it blue} contours indicates that the Halorun is sufficient for constructing a model that describes the selection bias well in the Fullrun.}
\label{fig:halovsfull}
\end{figure}

In section~\ref{sec:selection}, we develop the selection bias model based on \redmapper{} Halorun, where \redmapper{} is run fixing the cluster centers at the halo centers to avoid the ambiguity of associating galaxy clusters to dark matter halos. The \redmapper{} Halorun allows us to quantify the mass distribution of richness-selected galaxy clusters perfectly, which is important to quantify the selection bias. However, we will never be able to apply the same procedure on the data. Thus, it is crucial to understand the difference between \redmapper{} Halorun, and the actual \redmapper{} run (Fullrun), where \redmapper{} is run with the same setting as the run on real data. Fig.~\ref{fig:halovsfull} shows the comparison of the parameter constraints from the cluster abundance and two-point statistics measured from the Halorun and the Fullrun in one \textsc{Buzzard} realization. We note that we do not expect the parameter constraints from the Halorun and the Fullrun agree perfectly, since there is noise on the richness estimation in the Fullrun. Therefore, clusters whose richness is greater than 20, the lower richness cut in this analysis, in the Halorun have a different mass distribution from the clusters in the Fullrun. %
From Fig.~\ref{fig:halovsfull}, despite small differences, we can see both cosmological parameters ($\Omega_{\rm m}$ and $\sigma_8$), richness--mass relation parameters ($\rm{ln}\lambda_0$, $A_\lambda$, $\sigma_{\rm{intrinsic}}$, $B_\lambda$), and selection bias parameters ($b_{\rm{s0}}$, $b_{\rm{s1}}$), agree well. This indicates that the Halorun is sufficient for constructing a model that describes the selection bias well in the Fullrun. 

\section{Comparison of simulations to the data}
\label{app:datacomparison}
\begin{figure*}
\centering
\includegraphics[width=1.0\textwidth]{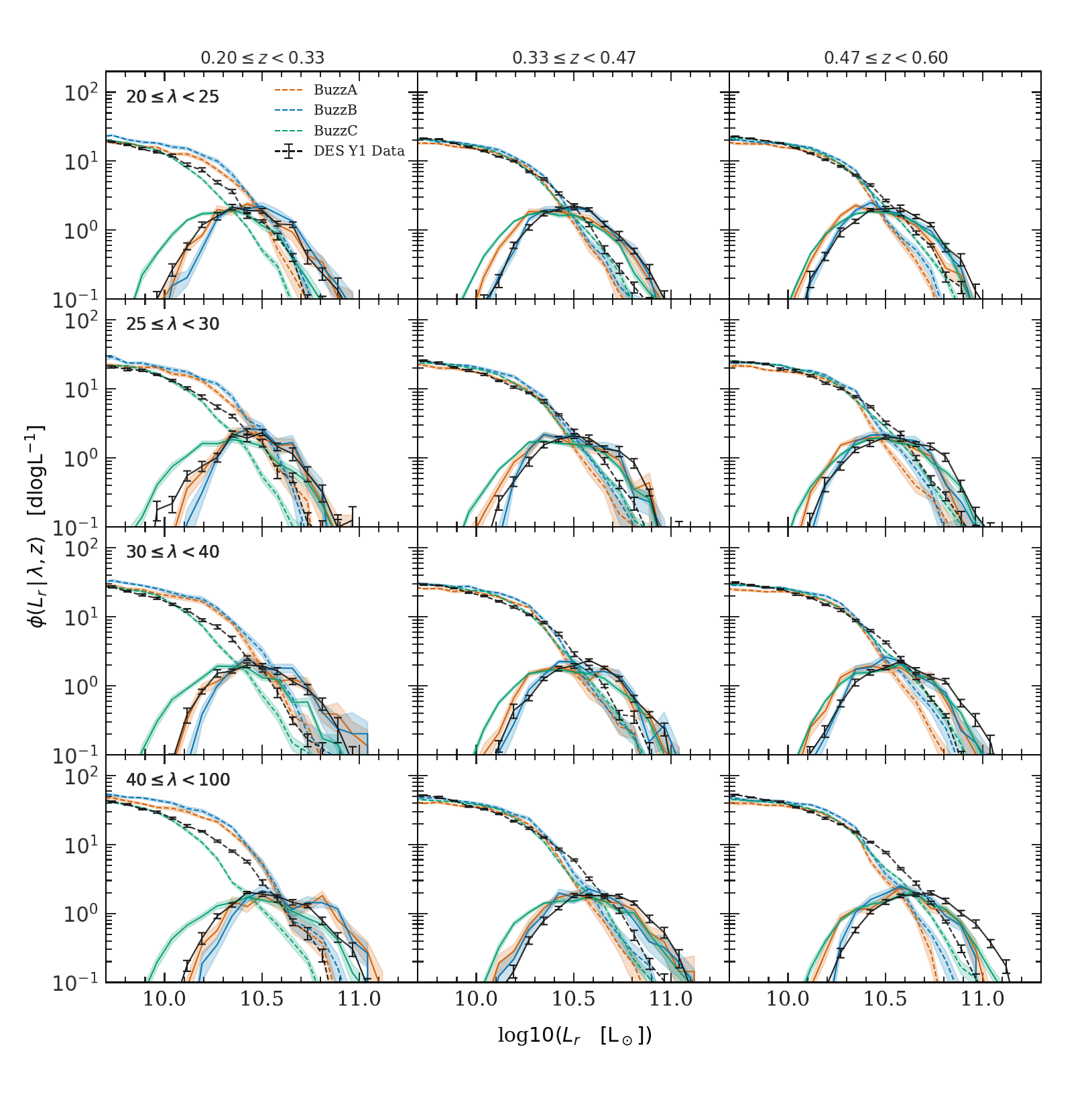}\hspace{-0.05\textwidth}
\caption{Conditional luminosity function as a function of redshift $z$ and richness $\lambda$. Richness increases from top to bottom and redshift increases from left to right. The dashed line and solid lines represent the luminosity function of satellite galaxies and central galaxies, respectively. The shaded region and error bar correspond to $1-\sigma$ uncertainties. The different colors correspond to different versions of \textsc{Buzzard} as summarized in Table~\ref{tab:sims}. For comparison, we overplot the measurement of the DES Y1 data. To account for the differences of richness--mass relations in simulations and data, we abundance match the simulations to the data. That is, for each richness bin, we consider the top $X $ per cent most massive halos in the simulation, where there are $X$ per cent of the clusters in the data that have richness greater than the given richness. In this plot, we focus on the relative brightness between centrals and satellites, an important quantity related to \redmapper{} performance. The range between the three simulations spans the the data. 
}
\label{fig:clf}
\end{figure*}

In this section, we compare the conditional luminosity function in the simulations to the DES Y1 data. The conditional luminosity function is a good tool to understand the relative brightness of satellites and central galaxies. Clearly, this quantity is closely related to the performance of the cluster finder. The measurement is done according to the method described in \cite{2019arXiv191001656T}. To account for the different richness--mass relations in simulations and data, we abundance match the simulations to the data. That is, for each richness bin, we consider the top $X $ per cent most massive halos in the simulation, where there are $X$ per cent of the clusters in the data that have richness greater than the given richness. Fig.~\ref{fig:clf} shows that while the conditional luminosity functions in the simulations do not match the data exactly, the range spanned by different simulations well covers the data.  

\section{Tinker vs Emulator}
\label{app:tinkeremulator}
\begin{figure}
\centering
\includegraphics[width=0.5\textwidth]{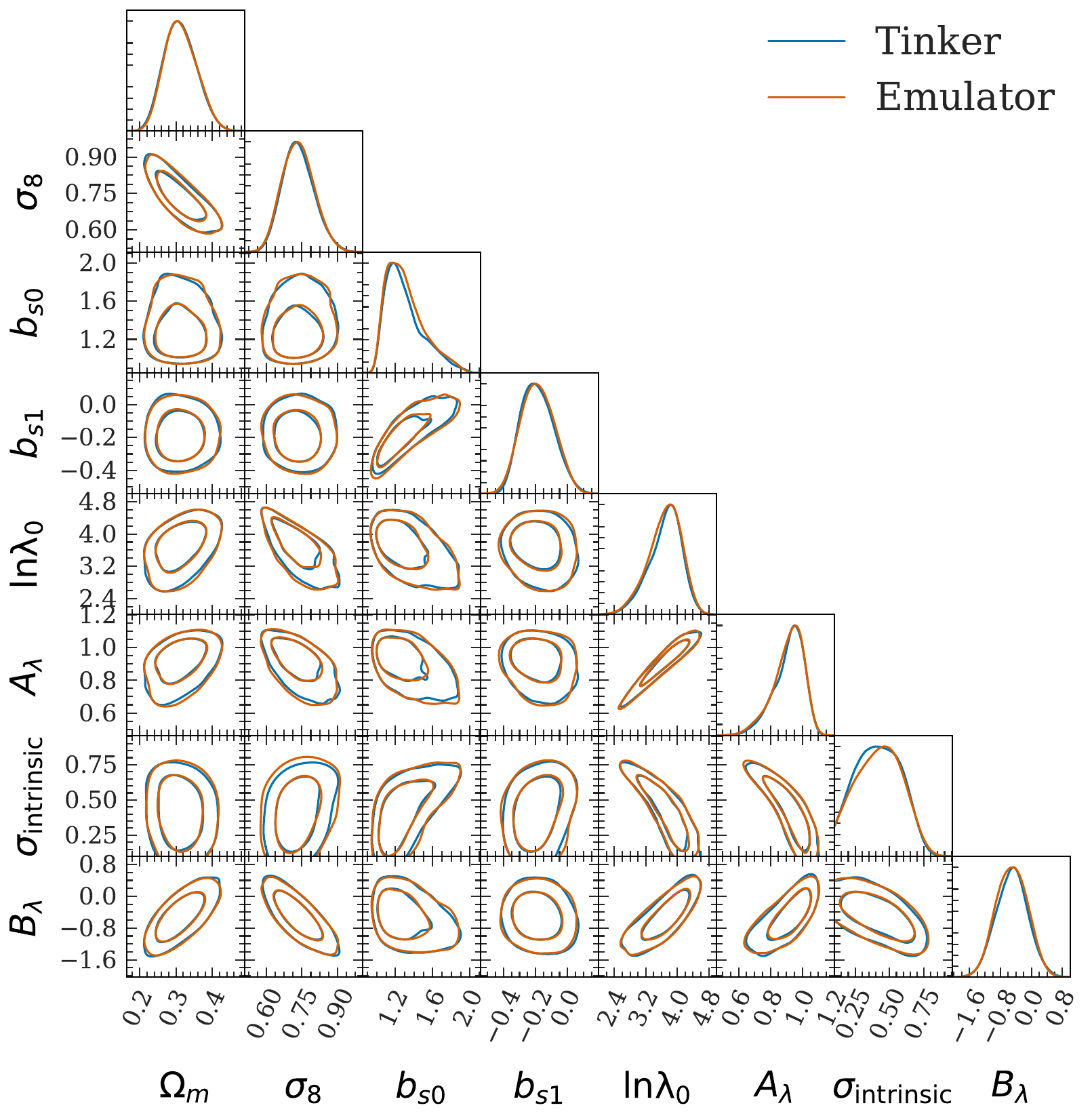}\hspace{-0.05\textwidth}
\caption{
Constraints on cosmological parameters ($\Omega_{\rm m}$ and $\sigma_8$), richness--mass relation parameters ($\rm{ln}\lambda_0$, $A_\lambda$, $\sigma_{\rm{intrinsic}}$, $B_\lambda$), and selection bias parameters ($b_{\rm{s0}}$, $b_{\rm{s1}}$); contours show the $68$ per cent and $95$ per cent confidence levels.  Both colors are analysis on data vectors generated from the same \textsc{Buzzard} realization and the same covariance matrix.  The {\it blue} lines represent the analysis using the Tinker halo mass function and the Tinker bias. The {\it orange} lines use the Aemulus emulators \citep{HMFemulator, biasemu}.  The agreement between the {\it orange} and {\it blue} contours indicates that the systematic uncertainties in the Tinker halo mass function and the Tinker bias are subdominant in this analysis. 
}
\label{fig:tinkervsemu}
\end{figure}

One potential systematic in this analysis is the use of the Tinker halo mass function and the Tinker bias \citep{Tinker10}, which are known to have $5-10$ per cent systematic uncertainties \citep{HMFemulator, biasemu}. We test whether these uncertainties are ignorable in our analysis by comparing the parameter constraints from one \textsc{Buzzard} realization estimated using the Tinker halo mass function and the Tinker bias with the constraints estimated using the halo mass function emulator \citep{HMFemulator} and the halo bias emulator \citep{biasemu}. Fig.~\ref{fig:tinkervsemu} shows the comparison of constraints on both cosmological parameters and nuisance parameters.  The constraints from the Tinker halo mass function and the Tinker bias are consistent with the constraints from the emulator,  indicating that the theory systematics from the Tinker halo mass function and the Tinker bias are subdominant in this analysis. We note that the above conclusion is only valid for a DES-Y1 like survey with cosmological parameters considered in this work. Additional tests are needed for applications on more sophisticated cosmological models and on future surveys with higher precision.

\section{Covariance matrix}
\label{app:covariance}
\begin{figure}
\centering
\includegraphics[width=0.5\textwidth]{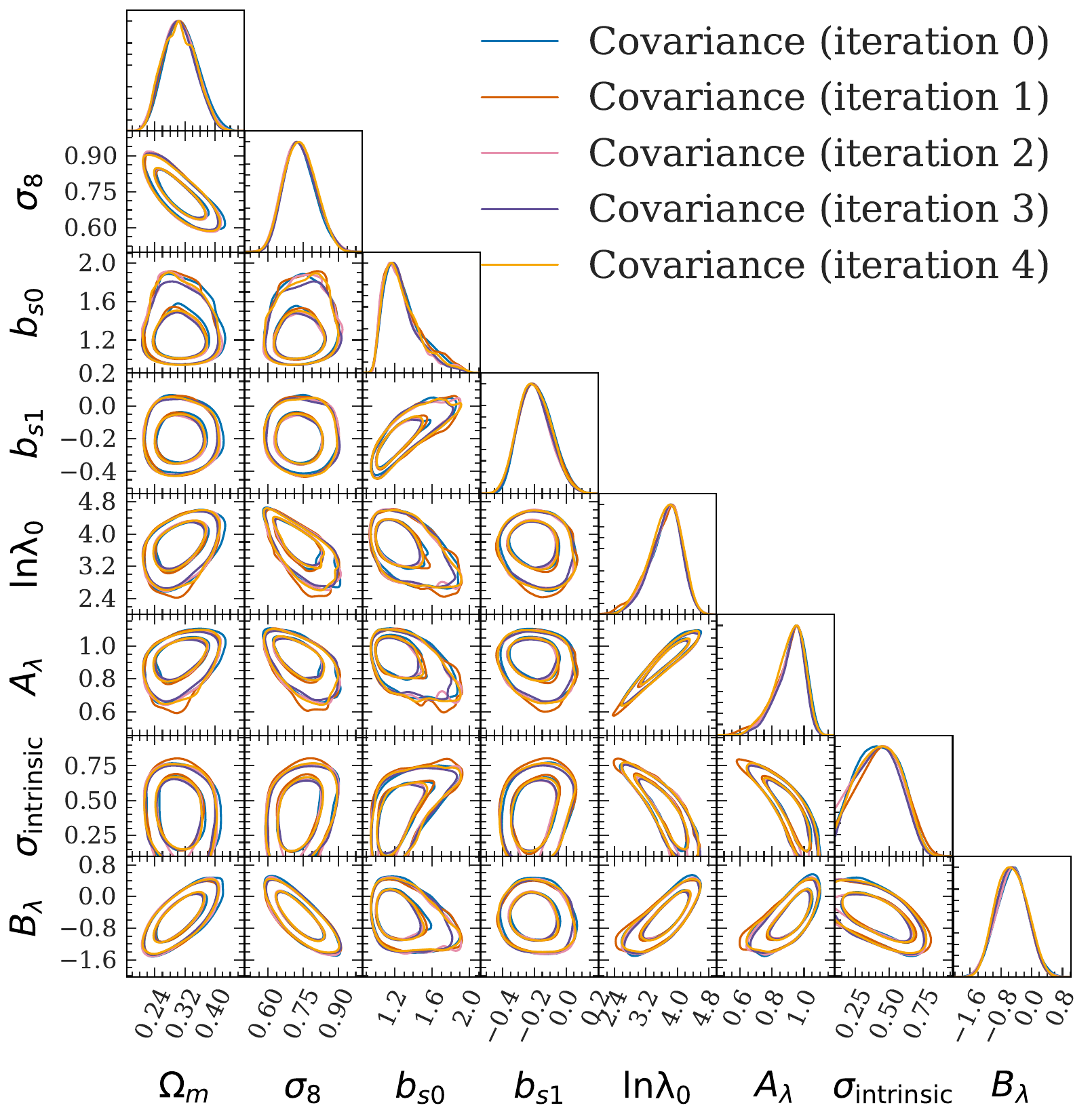}\hspace{-0.05\textwidth}
\caption{
 Constraints on cosmological parameters ($\Omega_{\rm m}$ and $\sigma_8$), richness--mass relation parameters ($\rm{ln}\lambda_0$, $A_\lambda$, $\sigma_{\rm{intrinsic}}$, $B_\lambda$), and selection bias parameters ($b_{\rm{s0}}$, $b_{\rm{s1}}$); contours show the $68$ per cent and $95$ per cent confidence levels.  The different colored contours show analysis on the same data vector with  different covariance matrices. The covariance matrix is generated with an iterative approach, where we update the richness--mass relation parameters used to generate the covariance matrix based on the constraints from the previous iteration. The agreement between subsequent iterations indicates that the parameter constraints are not sensitive to changes in the shot-noise estimation. }
\label{fig:covcomp}
\end{figure}

The cluster shot noise in the theory covariance matrix depends on the expected cluster abundance, which is sensitive to the richness--mass relation parameters. To determine the richness--mass relation parameters of a given cosmology, we adopt an iterative approach. Given a cosmology, we first decide a fiducial richness--mass relation parameters to generate a covariance matrix. This covariance matrix is then used to analyze the cluster abundance and the clustering of galaxy clusters to determine the richness--mass relation parameters. We repeat this process until convergence. Since there is noise in the shot noise estimation, it will be worrisome if the parameter constraints are sensitive to noise in the shot-noise estimation.  To test this, we run our analysis with the covariance matrix generated in each iterations of the above iterative process. Fig.~\ref{fig:covcomp} shows that the parameter constraints are not sensitive to noise in the shot-noise estimation. 

\section{Redshift dependent selection bias}
\label{app:bselz}

\begin{figure}
\centering
\includegraphics[width=0.5\textwidth]{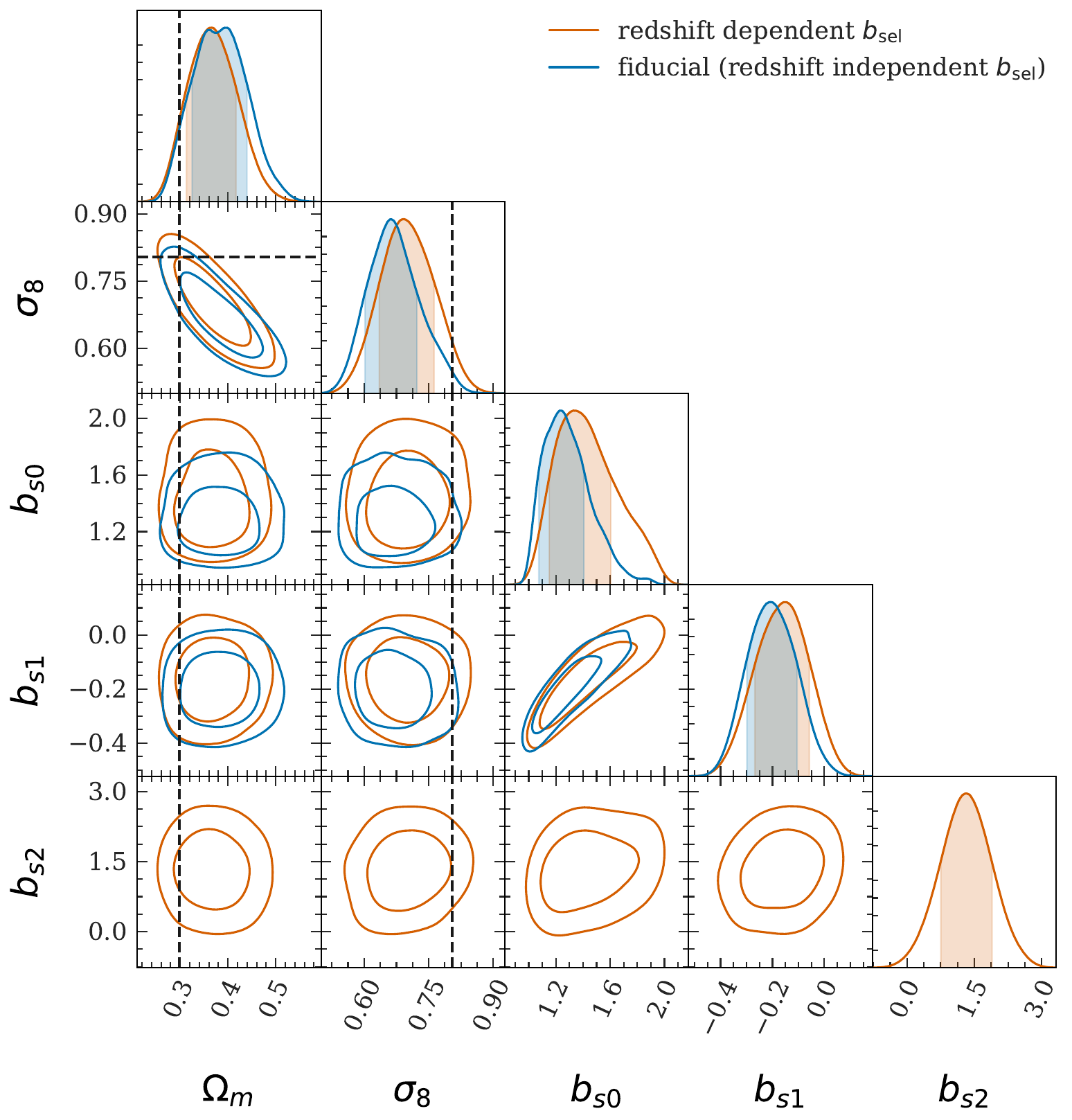}\hspace{-0.05\textwidth}
\caption{
 Constraints on cosmological parameters ($\Omega_{\rm m}$ and $\sigma_8$) and selection bias parameters ($b_{\rm{s0}}$, $b_{\rm{s1}}$, $b_{\rm{s2}}$).; contours show the $68$ per cent and $95$ per cent confidence levels.  $b_{\rm{s2}}$ is the redshift dependence of $b_{\rm{sel}}$ defined in equation \ref{eq:bselz}. The dashed lines denote the input cosmology of generating the simulation. The blue contours show the constraint assuming redshift independent $b_{\rm{sel}}$, and the orange contours show  the constraint assuming redshift dependent $b_{\rm{sel}}$. }
\label{fig:bselz}
\end{figure}

We consider the redshift-dependent $b_{\rm{sel}}$, defined as
\begin{equation}
\label{eq:bselz}
        b_{\rm{sel}}(M) = b_{s0}(M/M_{\rm{piv}})^{b_{s1}}\left(\frac{1+z}{1.45}\right)^{b_{s2}}, 
\end{equation}
where $b_{s2}$ describes the redshift dependency of $b_{\rm{sel}}$. We fit this model to the measured $b_{\rm{sel}}$ in the Halorun of each realization and find 5 out of 21 realizations show $b_{s2} \neq 0$ at $2$-$3\sigma$ significance. To test whether the possible redshift dependency of $b_{\rm{sel}}$ can bias our cosmological constraint, we rerun the analysis with the redshift-dependent selection bias model (equation~\ref{eq:bselz}) on realization 4a of \buzzardb{}, where we find the strongest detection of $b_{s2}\neq 0$ among all realizations. The result is shown in Fig.~\ref{fig:bselz}. We find that even in the most extreme case, including $b_{s2}$ into the posterior only shifts the contours of $\sigma_8$ and $\Omega_{\rm m}$ by $\approx 0.5 \sigma$.

\section{Comparison with galaxy--galaxy lensing and galaxy clustering} 
\label{app:2x2pt_4x2pt}
\begin{figure}
\label{fig:app:2x2pt_clustercomparison}
\includegraphics[width=0.4\textwidth]{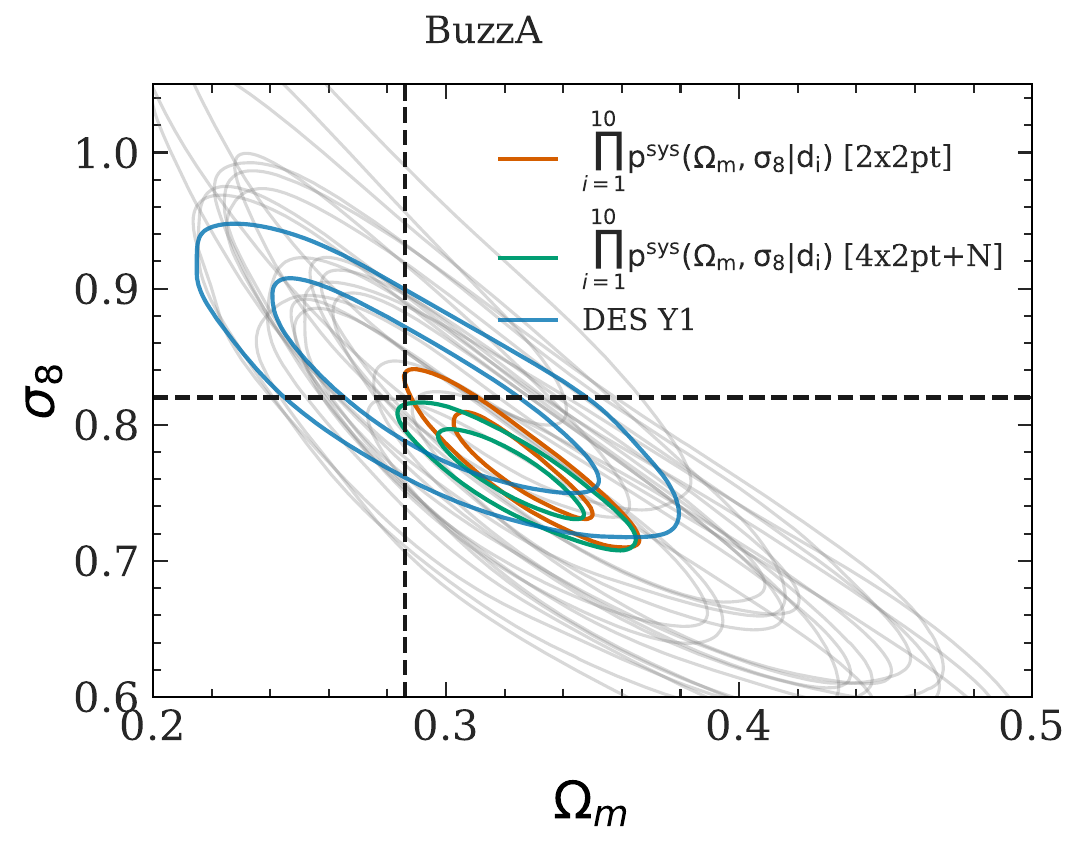}
\includegraphics[width=0.4\textwidth]{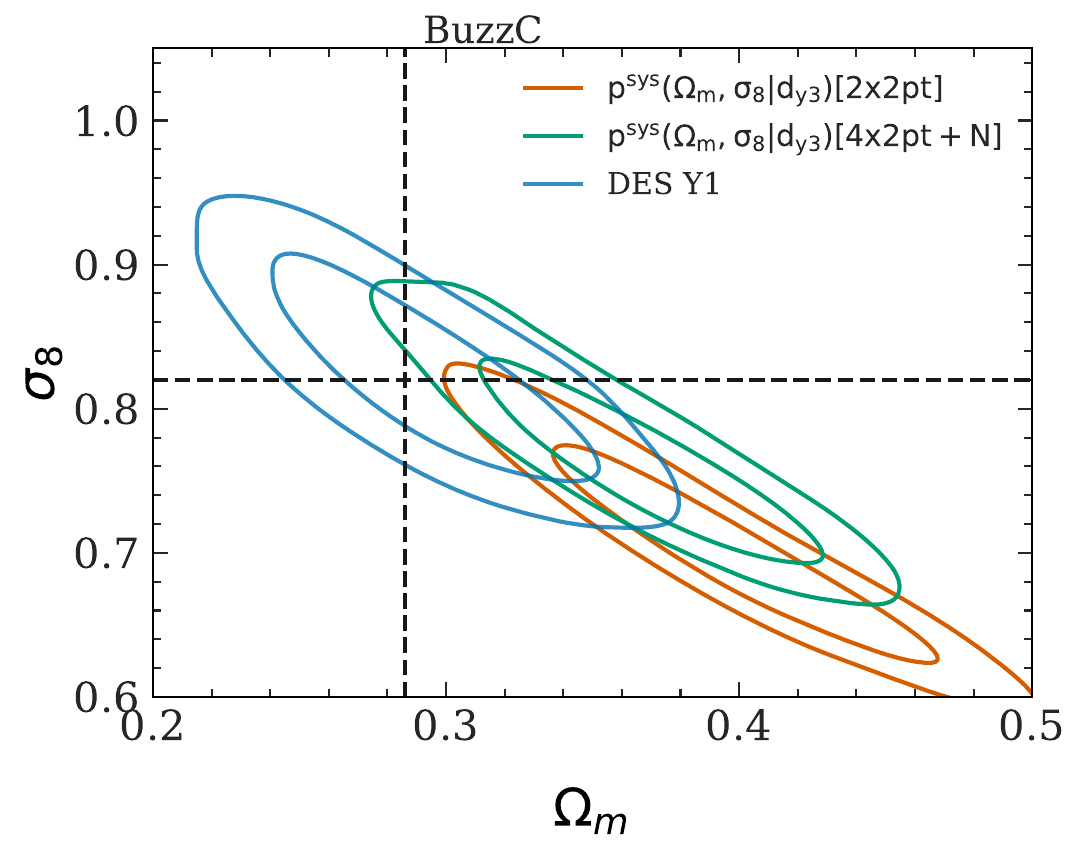}
\includegraphics[width=0.4\textwidth]{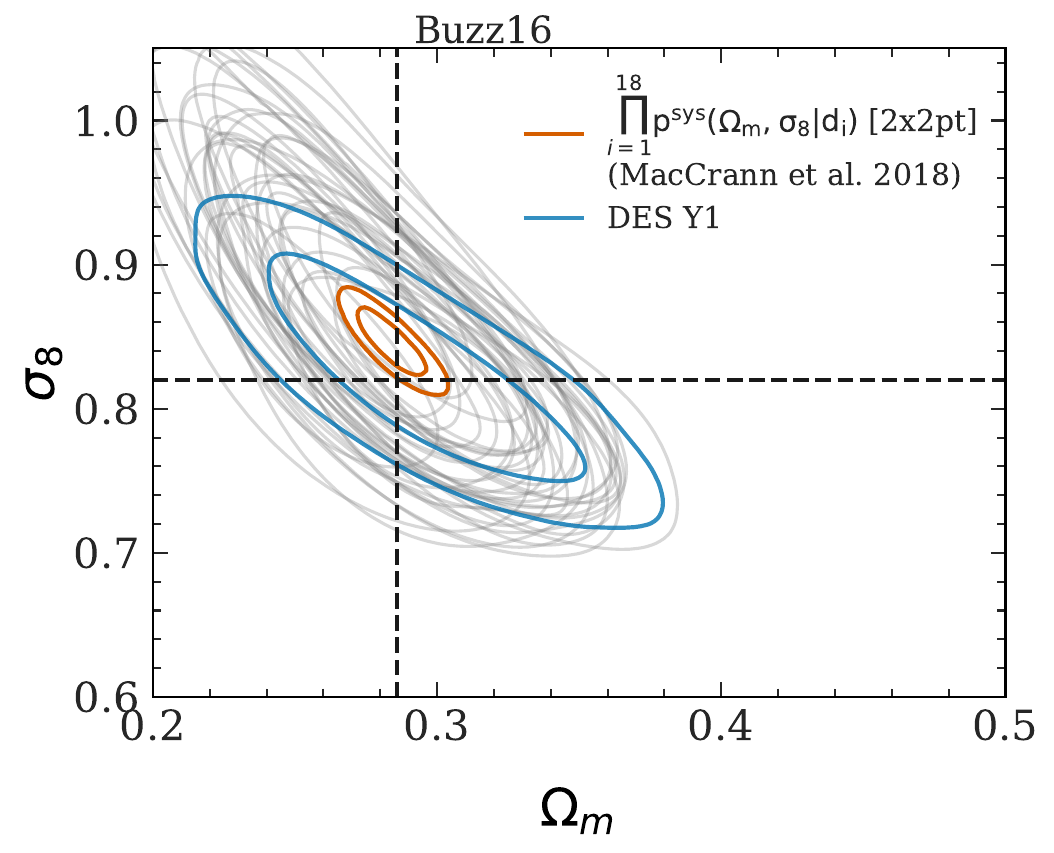}
\caption{
Constraints on $\Omega_{\rm{m}}$ and $\sigma_8$ from galaxy clustering and galaxy--galaxy lensing (2x2pt) measured in \buzzarda{} (top panel), \buzzardc{} (middle panel) and \textsc{Buzzard} v1.6 (bottom panel); contours show the $68$ per cent and $95$ per cent confidence levels.  In each panel, {\it gray} contours show the constraints from individual realizations and {\it orange} contours show the combination of these posteriors (equation~ \ref{eq:systematic}). As a comparison, green contours show the constraints from cluster number counts and four two-point correlation functions (\redmagic{} auto-correlations, \redmapper{}-\redmagic{} cross-correlations, \redmapper{} auto-correlations, and cluster lensing), same as the orange contours in Fig.~\ref{fig:countour_main}. 
The {\it blue} contours show the expected DES Y1 constraints shifted to center on the input cosmology of the simulation (same as the blue contours in Fig.~\ref{fig:countour_main}). The {\it black} dashed lines indicate the true cosmology, i.e. the input cosmology used to generate the simulation. The bottom panel shows the result presented in \citealt{Niall}, which validates the 2$\times$2pt pipeline on an older version of the \textsc{Buzzard} mocks (\textsc{Buzzard} v1.6).}
\end{figure}

To understand whether the $2$-$3\sigma$ systematics shown in Fig.~\ref{fig:countour_main} are due to flaws in the analysis pipeline or other sources, such as statistical fluctuations or flaws in simulations, we run an analysis combining galaxy-galaxy lensing and galaxy clustering (2$\times$2pt analysis) on each realizations of \buzzarda{} and \buzzardc{}. We combine the constraints on $\sigma_8$ and $\Omega_{\rm{m}}$ from analysis of each realization in the same way as we did in section ~\ref{sec:result}. The result is shown in Fig.~\ref{fig:app:2x2pt_clustercomparison}. We find that the 2x2pt analysis exhibits a similar bias in $\sigma_8$ and $\Omega_{\rm{m}}$ as the analysis  combining cluster counts and four two-point correlation functions. Further, in the bottom panel of Fig.~\ref{fig:app:2x2pt_clustercomparison}, we show the 2$\times$2pt analysis on an older version of the Buzzard mocks (Buzzard v1.6), presented in \cite{Niall}. The 2$\times$2pt analysis exhibits much less amounts of systematics in \textsc{Buzzard} v1.6 than in \buzzarda{} and \buzzardc{}. Thus, we believe that the $2$-$3\sigma$ systematics shown in Fig.~\ref{fig:countour_main} are not due to flaws in the analysis pipeline. More simulations are required to understand whether this is due to flaws in simulations of just statistical fluctuations. We leave this to future studies. 

\clearpage
\pagebreak

\end{document}